\documentclass[11pt, a4paper]{article}
\pdfoutput=1
\usepackage[utf8]{inputenc}
\usepackage[T1]{fontenc}
\usepackage{mlmodern}
\usepackage{textcomp}
\usepackage[american]{babel}
\usepackage{microtype}
\usepackage{csquotes}

\usepackage{eurosym}
\usepackage{xspace}
\usepackage[nottoc]{tocbibind}
\usepackage{authblk}
\usepackage{fancyhdr}
\usepackage{totpages}

\usepackage{setspace}
\usepackage{titlesec}
\usepackage{changepage}

\usepackage{graphicx}
\usepackage{subcaption}
\usepackage{float}
\usepackage{makecell}
\usepackage{multirow}
\usepackage{multicol}
\usepackage{enumitem}
\usepackage{marginnote}
\usepackage[multiple]{footmisc}

\usepackage[usenames, dvipsnames, svgnames, table]{xcolor}

\usepackage[
	backend=bibtex8, citestyle=numeric-comp,
	sorting=none, sortcase=false, sortcites=true,
	giveninits=true, maxnames=99
]{biblatex}

\usepackage{amsmath, amsfonts, amssymb}
\usepackage{mathtools}
\usepackage{mathrsfs}
\usepackage{cancel}
\usepackage{braket}
\usepackage{tensor}
\usepackage{slashed}

\usepackage[
	pdftex, breaklinks=true, linktocpage, colorlinks=true,
	urlcolor=RoyalBlue, linkcolor=RoyalBlue, citecolor=BrickRed
]{hyperref}
\usepackage[all]{hypcap}

\DeclareUnicodeCharacter{00A0}{~}
\DeclareUnicodeCharacter{202F}{~}

\newcommand{\email}[1]{\href{mailto:#1}{\nolinkurl{#1}}}
\newcommand{\emailfoot}[1]{\thanks{\email{#1}}}

\newcommand{\doi}[1]{\href{http://dx.doi.org/#1}{\nolinkurl{#1}}}
\newcommand{\urlhttp}[1]{\href{http://#1}{\nolinkurl{#1}}}

\newcounter{draftcommentcnt}
\NewDocumentCommand{\draftcomment}{s O{red} m}{%
	\def\margnote{\IfBooleanTF{#1}{\marginnote}{\marginpar}}%
	\stepcounter{draftcommentcnt}%
	\textcolor{#2}{#3}%
	\margnote{\textcolor{#2}{$\Leftarrow$ \arabic{draftcommentcnt}}}%
}

\graphicspath{{./images/}}
\numberwithin{equation}{section}

\usepackage[sort&compress, english]{cleveref}

\usepackage[heightrounded,
	top=3.5cm, bottom=3cm, left=3.5cm, right=3.5cm,
]{geometry}

\fancypagestyle{pageof}{
	\fancyhf{}

	\fancyfoot[C]{-- Page \thepage\ of \pageref*{TotPages} --}
}

\newcommand{\preprint}[0]{XXX}
\fancypagestyle{preprint}{
	\fancyhf{}

	\fancyhead[R]{\small\textsf{\preprint}}
}

\newcommand{\e}[0]{\mathrm{e}}
\newcommand{\I}[0]{\mathrm{i}}
\newcommand{\N}[0]{\mathbb{N}}
\newcommand{\Z}[0]{\mathbb{Z}}

\newcommand{\R}[0]{\mathbb{R}}
\newcommand{\C}[0]{\mathbb{C}}
\newcommand{\dd}[0]{\mathrm{d}}

\newcommand{\pd}[0]{\partial}
\newcommand{\mc}[1]{{\mathcal{#1}}}

\renewcommand{\vec}[1]{\boldsymbol{#1}}

\DeclareMathOperator{\tr}{tr}
\DeclareMathOperator{\diag}{diag}

\newcommand{\mean}[1]{\langle #1 \rangle}
\newcommand{\Mean}[1]{\left \langle #1 \right \rangle}
\newcommand{\abs}[1]{{|#1|}}

\renewcommand{\bra}[2][]{\mathinner{{}_{#1}\hspace{-2pt}\langle #2 |}}

\newcommand{\com}[2]{[ #1, #2 ]}

\newcommand{\group}[1]{\mathrm{#1}}

\newcommand{\conj}[1]{{#1^*}}
\newcommand{\adj}[1]{{#1^\dagger}}

\renewcommand{\=}[1]{\stackrel{\text{#1}}{=}}

\input{arxiv.def}

\hypersetup{
	pdftitle={Gravitational action for a massive Majorana fermion in 2d quantum gravity},
	pdfauthor={Corinne de Lacroix, Harold Erbin, Vincent Lahoche}
}

\title{Gravitational action for a massive Majorana fermion in $2d$ quantum gravity}

\author[1]{Corinne de Lacroix\emailfoot{corinne.delacroix@mailo.com}}
\author[2,3,4]{Harold Erbin\emailfoot{erbin@mit.edu}}
\author[2]{Vincent Lahoche\emailfoot{vincent.lahoche@cea.fr}}

\affil[1]{%
	LPT, Département de physique de l'ENS, École normale supérieure
	\protect\\
	UPMC Univ. Paris 06, CNRS, PSL Research University
	\protect\\
	75005 Paris, France
}

\affil[2]{%
	Université Paris-Saclay, CEA, LIST, F-91120 Palaiseau, France
}

\affil[3]{%
	Center for Theoretical Physics, Massachusetts Institute of Technology
	\protect\\
	Cambridge, MA 02139, USA
}

\affil[4]{%
	NSF AI Institute for Artificial Intelligence and Fundamental Interactions
}

\DeclareMathOperator{\grad}{\nabla}
\newcommand{\anticom}[2]{\{ #1, #2 \}}
\newcommand{\bracket}[2]{\mathinner{\langle #1 | #2 \rangle}}

\begin{document}

\maketitle

\begin{abstract}
We compute the gravitational action of a free massive Majorana fermion coupled to two-dimensional gravity on compact Riemann surfaces of arbitrary genus.
The structure is similar to the case of the massive scalar.
The small-mass expansion of the gravitational yields the Liouville action at zeroth order, and we can identify the Mabuchi action at first order.
While the massive Majorana action is a conformal deformation of the massless Majorana CFT, we find an action different from the one given by the David--Distler--Kawai (DDK) ansatz.
\end{abstract}

\newpage

\hrule
\pdfbookmark[1]{\contentsname}{toc}
\tableofcontents
\bigskip
\hrule

\newpage

\section{Introduction}

The functional integral is the most direct approach to quantum gravity coupled to matter.
However, this strategy is difficult to carry out in four dimensions: first, because of the computational complexity of the functional integral, second, because gravity is non-renormalizable.

In such circumstances, a natural exercise is to study the corresponding theory in a lower number of dimensions, in particular, in two dimensions.
In this case, computations can often be carried out explicitly and one can gain intuition for approaching and understanding the higher-dimensional problem.
This paper will thus be concerned about two-dimensional quantum gravity, which is particularly simple because the Einstein--Hilbert action is trivial and only the matter and the cosmological constant contribute to the dynamics of gravity.

Since the metric in two dimensions contains only one degree of freedom, the standard approach consists in splitting the metric into a conformal factor, called the Liouville field $\sigma$, and a fixed background metric $\hat g_{\mu\nu}$.
Then, the dependence in the Liouville mode of the functional integral can be factored from the matter by introducing a Wess--Zumino--Witten term -- called the  gravitational action -- for the Liouville field.

In his seminal paper~\cite{Polyakov:1981:QuantumGeometryBosonic}, Polyakov found that the gravitational action for a scalar field is given by the famous Liouville action.
In fact, this statement holds for any conformal field theory (CFT) by relating the action to the conformal anomaly on a curved manifold.
The properties of this theory have been largely investigated, and the reader is refereed to~\cite{Curtright:1982:ConformallyInvariantQuantization, DHoker:1982:ClassicalQuantalLiouville, Braaten:1983:ExactOperatorSolution, Gervais:1985:Nonstandard2DCritical, David:1988:ConformalFieldTheories, Distler:1989:ConformalFieldTheory, Seiberg:1990:NotesQuantumLiouville, Dorn:1994:TwoThreepointFunctions, Zamolodchikov:1996:StructureConstantsConformal, Harlow:2011:AnalyticContinuationLiouville, Bilal:2015:2DQuantumGravity, Leduc:2016:2DQuantumGravity, Bautista:2019:QuantumGravityTimelike,Bautista:2020:BRSTCohomologyTimelike} for a selected set of references and to the reviews~\cite{Ginsparg:1993:Lectures2DGravity, Teschner:2001:LiouvilleTheoryRevisited, Nakayama:2004:LiouvilleFieldTheory, Zamolodchikov:2007:LecturesLiouvilleTheory, Ribault:2022:ConformalFieldTheory} for additional details.

For a model to be faithful to our four-dimensional world, it should contain non-conformal matter.\footnotemark{}
\footnotetext{%
	Some aspects of massive matter coupled to classical two-dimensional gravity have been recently discussed in~\cite{deLacroix:2020:ShortNoteDynamics}.
}%
Surprisingly, this topic has been mostly ignored in the literature and only the last decade saw the beginning of an investigation from first-principles based on the heat kernel.
The first terms of a development in the mass of the gravitational action for a massive scalar field (with and without non-minimal coupling to gravity) on a compact Riemann surface have been computed in~\cite{Ferrari:2011:RandomGeometryQuantum, Ferrari:2012:GravitationalActionsTwo, Ferrari:2014:FQHECurvedBackgrounds}.
Besides the Liouville action, two other functionals well-known from the mathematicians appear in the development: the Mabuchi\footnotemark{} and the Aubin--Yau functionals~\cite{Mabuchi:1986:KenergyMapsIntegrating, Phong:2008:LecturesStabilityConstant}.
\footnotetext{%
	The string susceptibility and spectrum of this action have been discussed in~\cite{Bilal:2014:2DQuantumGravity, deLacroix:2016:MabuchiSpectrumMinisuperspace, deLacroix:2018:MinisuperspaceComputationMabuchi}.
	A rigorous mathematical construction based on a generalization of the Gaussian multiplicative chaos has been given in~\cite{Lacoin:2018:PathIntegralQuantum}.
}%
Finally, the gravitational action has later been computed completely in the case of manifolds with and without boundaries~\cite{Bilal:2017:2DQuantumGravity, Bilal:2017:2DGravitationalMabuchi}.

The historical approach to the coupling of non-conformal matter to $2d$ gravity follows the David--Distler--Kawai (DDK) construction~\cite{David:1988:ConformalFieldTheories, Distler:1989:ConformalFieldTheory}.
It provides an ansatz for the gravitational action in two situations: 1) when the matter is a CFT deformed by a set of primary operators, 2) when changing to the free field measure for the Liouville mode.
The second case has received strong support, both directly from an explicit derivation~\cite{Mavromatos:1989:RegularizingFunctionalIntegral, DHoker:1990:2DQuantumGravity, DHoker:1991:EquivalenceLiouvilleTheory}, and indirectly from the rigorous constructions of the Liouville CFT using the conformal bootstrap~\cite{Teschner:2001:LiouvilleTheoryRevisited, Ribault:2022:ConformalFieldTheory, Ribault:2015:LiouvilleTheoryCentral} or the Gaussian multiplicative chaos~\cite{Rhodes:2016:LectureNotesGaussian, Kupiainen:2016:ConstructiveLiouvilleConformal}.
On the other hand, it is known that the first case presents some problems.
In particular, the $\beta$-functions associated with the matter coupling constants do not vanish, and the theory is not conformally invariant at the quantum level~\cite{Schmidhuber:1993:ExactlyMarginalOperators, Ambjorn:1994:2DQuantumGravity}.
Moreover, the functional integral of the Liouville field cannot be factorized from the matter functional integral.

The goal of this paper is to compute the gravitational action of a massive Majorana fermion on a compact Riemann surface of arbitrary genus (without boundary), developing the computations performed in~\cite{Bilal:2021:EffectiveGravitationalAction} on the sphere (see also~\cite{Erbin:2021:GravitationalActionMassive} for earlier computations and alternative methods).
On flat space, a massive Majorana fermion in two dimensions is equivalent to the Ising model at finite temperature (or massive Ising model)~\cite{Zuber:1977:QuantumFieldTheory, Bander:1977:QuantumfieldtheoryCalculationTwodimensional}.
The latter is a conformal deformation of the Ising model, which is a CFT with central charge $c = 1/2$~\cite{DiFrancesco:1999:ConformalFieldTheory}.

It is interesting to study the coupling of this model to gravity for two reasons.
First, it is the simplest model after the massive scalar field, since it is also a free theory.
Thus, it is expected that the gravitational action can be explicitly computed.
This point would be helpful to provide more insights on the possible sources for the Mabuchi action.
In a second time, one can compare this expression with the ansatz provided by the DDK construction.
Ultimately, we find that both actions \emph{do not agree}, which calls for a deeper study of the DDK ansatz.

\paragraph{Outline}

In \Cref{sec:majorana-field-theory}, we introduce the field theory of a massive Majorana fermion on a curved space.
We discuss in length how to define the functional integral of a Majorana field in a general basis.
This yields an expression for the effective action as a functional determinant.
\Cref{sec:spectral-analysis} is devoted to defining the tools needed for the spectral analysis.
In particular, we define the spectral functions which appear in the computations (Green functions, zeta function, heat kernel) and derive their conformal variations.
We also explain how to handle zero-modes properly.
The gravitational action for a Majorana fermion coupled to $2d$ gravity on a Riemann surface of arbitrary genus is derived in \Cref{sec:gravitational-action}, see \eqref{expression-Sgrav-final}, which is the main result of our paper.
We also comment on the small mass expansion and show that the first correction contains the Mabuchi action.
\Cref{sec:conventions} summarizes our conventions, while \Cref{sec:2d-fermions} gathers formulas on two-dimensional Euclidean fermions and gamma matrices.
In \Cref{sec:ddk-ansatz}, we recall the DDK construction, describe its possible problems and discuss them, in view of our results.
Finally, in \Cref{app:proof-identities-Kzeta}, we give some detailed computations of identities that were used in \ref{sec:gravitational-action}.

\bigskip

\textbf{Note added:} \emph{While we were finalizing our manuscript, the paper~\cite{Namuduri:2023:EffectiveGravitationalAction} appeared and obtained results similar with this paper and the earlier draft~\cite{Erbin:2021:GravitationalActionMassive} (written prior to~\cite{Bilal:2021:EffectiveGravitationalAction} and which already computed the zero-modes, projectors and finite variations) for arbitrary genus.}

\section{Majorana fermion field theory}
\label{sec:majorana-field-theory}

In this section, we present the action and the associated functional integral of a two-dimensional massive Majorana fermion coupled to gravity in Euclidean signature.
We consider a compact Riemann surface without boundaries.
Conventions are given in \Cref{sec:conventions}, and general properties of gamma matrices and spinors in two dimensions are given in \cref{sec:2d-fermions}.

\subsection{Classical action}

The action for a two-dimensional Majorana fermion $\Psi$ coupled to gravity is given by~\cite{Zuber:1977:QuantumFieldTheory, Polyakov:1981:QuantumGeometryFermionic}
\begin{equation}
	S_{\text{m}}[g, \Psi]
		= \frac{1}{4\pi} \int \dd^2 x\, \sqrt{g} \,
			\bar\Psi (\I \slashed\grad + m \gamma_*) \Psi,
\end{equation}
where the Dirac adjoint reads $\bar\Psi = \adj{\Psi}$ and $m$ is the mass.\footnotemark{}
\footnotetext{%
	The Dirac conjugation is sometimes defined by $\bar\Psi = \adj{\Psi} \gamma^0$~\cites[sec.~5.3.2]{DiFrancesco:1999:ConformalFieldTheory}[sec.~9.7]{Mussardo:2009:StatisticalFieldTheory}, but the above object does not transform appropriately under Lorentz transformations (see \cref{sec:2d-fermions}).
}%
The Dirac operator $\slashed\grad$ is defined as (see \Cref{sec:conventions} for more details):
\begin{equation}
	\slashed\grad
		:= \gamma^\mu \grad_\mu,
	\qquad
	\grad_\mu \Psi
		:= \pd_\mu \Psi + \Gamma_\mu \Psi,
\end{equation}
where the connection reads:
\begin{equation}
	\Gamma_\mu
		:= - \frac{\I}{4} \, \omega_\mu \gamma_*,
	\qquad
	\omega_\mu
		:= \omega_{\mu a b} \epsilon^{ab},
\end{equation}
with $\omega_{\mu a b}$ is the spin connection, $\epsilon_{01} = 1$ is the antisymmetric Levi--Civita symbol, and $\gamma_* = \I \gamma^0 \gamma^1$.
Curved indices are denoted with Green letters ($\mu, \nu$, etc.), and indices of the local flat frame by Latin letters ($a, b$, etc.).
One has to introduce the chirality matrix $\gamma_*$ in the mass term, since the standard bilinear $\bar{\Psi} \Psi$ vanishes (see \cref{sec:2d-fermions}).
Finally, let us mention that $\I \slashed\grad$ is Hermitian with the usual inner-product, defined below in \eqref{eq:inner-product}, since $\gamma_*$ is Hermitian and the complex conjugation of the factor $\I$ compensates the sign arising from the integration by parts.

Even though the connection term vanishes due to the flip relation \eqref{eq:flip-majorana}~\cite{Blumenhagen:2014:BasicConceptsString}, one needs to keep it when defining the functional integral in order to work with a covariant object.
In particular, in the Weyl basis, writing $\Psi = (\bar\psi, \psi)$ this action reads
\begin{equation}
	S[g, \Psi] = \frac{1}{4\pi} \int \dd^2 z\, \sqrt{g}\, (
		\psi \bar\pd \psi
		- \bar\psi \pd \bar\psi
		+ 2 m \, \psi \bar\psi)
\end{equation}
where $\pd = \frac{1}{2} (\pd_0 - i \pd_1)$ and $\bar\pd = \frac{1}{2} (\pd_0 + i \pd_1)$. This is the standard form for the massive Ising model~\cite[sec.~9.2.2]{Itzykson:1991:StatisticalFieldTheory-2}.

\subsection{Functional integral}

The partition function of the theory is given by the functional integral
\begin{equation}
	Z = \int \mc D g \mc D \Psi \, \e^{-S[g, \Psi] - S_{\mu}[g]}
		:= \int \mc D g \, Z[g],
\end{equation}
where $S_\mu[g] = \mu \int \dd^2 x \sqrt g = \mu A$ is the cosmological constant action.
In this paper, we will not be interested in the integration over the metrics but restrict ourselves to the matter part.
For this reason, we also do not take into account the Einstein--Hilbert action, which in two dimensions is just a constant proportional to the Euler number of the Riemann surface.
$Z[g]$ can be further decomposed as
\begin{equation}
	Z[g] = \e^{-S_\mu[g]} \int \mc D \Psi \, \e^{-S[g, \Psi]}
		:= \e^{-S_\mu[g]} \, Z_m[g].
\end{equation}
The effective action $S_{\text{eff}}[g]$ is defined by
\begin{equation}
	Z_m[g] := \e^{-S_{\text{eff}}[g]}.
\end{equation}
Computing $S_{\text{eff}}[g]$ directly is very challenging, and it is simpler to obtain the gravitational action
\begin{equation}
	\label{eq:def-gravitationnal-action}
	S_{\text{grav}}[\hat g, g]
		:= S_{\text{eff}}[g] - S_{\text{eff}}[\hat g]
		= - \ln \frac{Z_m[g]}{Z_m[\hat g]},
\end{equation}
where $g$ and $\hat g$ are two different metrics.
One then has
\begin{equation}
	Z[g] = \e^{-S_\mu[g]} \, \e^{-S_{\text{grav}}[\hat g, g]} \, Z_m[\hat g]
\end{equation}
and the complete action reads
\begin{equation}
	\label{eq:cg-action}
	S_{\text{cg}}[\hat g, g, \Psi]
		= S_{\text{grav}}[\hat g, g] + S_m[\hat g, \Psi].
\end{equation}

Since the action is quadratic, the path integral is Gaussian and reduces to the determinant of the kinetic operator
\begin{equation}
	\label{eq:operator-D}
	D = \I \slashed \nabla + m \gamma^*
\end{equation}
to some power.
The operator $D$ is Hermitian since $\I \slashed\nabla$ and $\gamma_*$ are Hermitian.
Moreover, it is possible to rewrite the determinant in terms of the square of the operator $D$
\begin{equation}
	\label{eq:operator-D2}
	D^2 = - \Delta + \frac{R}{4} + m^2,
\end{equation}
where $\Delta$ is the spinor Laplacian:
\begin{equation}
	\Delta \Psi
		:= g^{\mu\nu} \grad_\mu \grad_\nu \Psi.
\end{equation}
Note that $D^2$ is diagonal only in the Weyl basis \eqref{eq:weyl-basis}, where $\gamma_* = \diag(1, -1)$.
The rest of this section is devoted to the derivation of this result, which is obtained by diagonalizing (formally) the kinetic operator.

\subsubsection{Mode expansion}

There is no solution to the eigenvalue equation
\begin{equation}
	D \Psi
		= (\I \slashed\grad + m \gamma_*) \Psi
		\={?} \lambda \Psi
\end{equation}
with $\lambda \in \R$ (since $D$ is Hermitian) and such that $\Psi$ satisfies the Majorana condition:
\begin{equation}
	\conj{\Psi}
		= C \Psi,
\end{equation}
with $C$ the charge conjugation matrix defined in \eqref{eq:charge-conj-def}.
Indeed, by taking the conjugate of the equation and inserting the Majorana condition, one finds
\begin{equation}
	D \Psi
		= (\I \slashed\grad + m \gamma_*) \Psi
		= - \conj{\lambda} \Psi.
\end{equation}
This problem can be solved by looking for complex eigenvectors to be decomposed into their real and imaginary parts (under the Majorana conjugation).
Hence, we are looking for complex eigenfunctions $\Psi_n \in \C$ of $D$ with real eigenvalues $\lambda_n$ ($n \in \Z$)~\cite[sec.~13.3]{Wipf:2016:IntroductionSupersymmetry}:\footnotemark{}
\footnotetext{%
	Note that we could work with real eigenmodes by inserting the matrix $\gamma_*$ on the RHS.
	However, this complicates all expressions since this matrix would appear in the definition of the inner-product, Green functions, etc.
}
\begin{equation}
	D \Psi_n
		= (\I \slashed\grad + m \gamma_*) \Psi_n
		= \lambda_n \, \Psi_n,
	\qquad
	\lambda_n \in \R.
\end{equation}

The inner-product between two spinors $\psi_1$ and $\psi_2$ is defined as:
\begin{equation}
	\label{eq:inner-product}
		\bracket{\psi_1}{\psi_2}
		:= \frac{1}{2\pi} \int \dd^2 x \sqrt{g} \,
		\adj{\psi_1(x)} \psi_2(x).
\end{equation}
Note that it vanishes for anti-commuting Majorana spinors (for example, in the Majorana basis: $\adj{\psi} \psi = \psi^t \psi = 0$); however, this is not a problem since the eigenmodes are commuting and complex.\footnotemark{}
\footnotetext{%
	It would be necessary to add $\gamma_*$ in the definition of the inner-product if it also appears in the RHS of the eigenvalue equation.
	But, as pointed in the previous footnote, this makes all expressions much more complicated.
	One particular problem is that $D$ is not self-adjoint for this product, instead: $\bracket{\psi_1}{D \psi_2} = \bracket{\widetilde D \psi_1}{\psi_2}$, where $\widetilde D := \gamma_* D \gamma_*$.
}%
The eigenfunctions form a complete set and are taken to be orthonormal:\footnotemark{}
\footnotetext{%
	In fact, modes with $\lambda_n = m$ (where $m$ is the mass) are degenerate and generically not orthonormal, see \Cref{sec:zero-modes}.
	However, ignoring this subtlety does not change the computation in general and the fact that zero-modes are not orthonormal will be taken care of when needed.
}%
\begin{equation}
	\label{eq:mode-normalization-first}
	\bracket{\Psi_m}{\Psi_n}
		= \delta_{mn}.
\end{equation}
By computing the inner-product with an insertion of $D$, we can easily check that the eigenvalues are real:
\begin{equation}
\begin{aligned}
		\bracket{\Psi_n}{D \Psi_n}
			&
			= \lambda_n
		\\ &
			= \bracket{D \Psi_n}{\Psi_n}
			= \conj{\lambda_n}.
	\end{aligned}
\end{equation}

One can check that if $\lambda_n$ is the eigenvalue associated to $\psi_n$, then $- \lambda_n$ is the eigenvalue of $C^{-1} \conj{\Psi_n}$ for $n \neq 0$:
\begin{equation}
	D (C^{-1} \conj{\Psi_n})
		= - \lambda_n \, (C^{-1} \conj{\Psi_n}).
\end{equation}
As a consequence, we define
\begin{equation}
	\label{eq:mode-Psi-n-neg}
	\forall n \in \N^*:
	\qquad
	\Psi_{-n}
		:= C^{-1} \conj{\Psi_n},
	\qquad
	\lambda_{-n}
		:= - \lambda_n.
\end{equation}

The Majorana field is expanded on the modes $\Psi_n$ as:
\begin{equation}
	\Psi
		:= \sum_{n \ge 0} (a_n \Psi_n + a_{-n} \Psi_{-n}),
\end{equation}
where the $a_n$ are complex Grassmann variables and satisfy:
\begin{equation}
	\label{eq:an-dagger}
	a_{-n}
		= \adj{a_n}.
\end{equation}
Note that in this decomposition the coefficients are taken to be Grassmann numbers while the eigenfunctions are commuting functions.
As a consequence, the normalization \eqref{eq:mode-normalization-first} would be non-trivial even without inserting $\gamma_*$.
The coefficient $a_n$ can be recovered by taking the inner-product with $\Psi_n$:
\begin{equation}
	a_n
		= \bracket{\Psi_n}{\Psi}.
\end{equation}

The Dirac conjugate is:
\begin{equation}
	\bar\Psi
		= \sum_{n \ge 0} (a_n \Psi_n^t C + a_{-n} \Psi_{-n}^t C)
		= \sum_{n \ge 0} \big( a_n \bar\Psi_{-n} + a_{-n} \bar\Psi_n \big),
\end{equation}
using that $\bar\Psi = \bar\Psi^c = \Psi^t C$ since $\adj{(C^{-1})} = \adj{(\conj C)} = C^t = C$, and the second equality follows from \eqref{eq:mode-Psi-n-neg}.
Since $\bar\Psi = \adj{\Psi}$, one can also recover the expression \eqref{eq:an-dagger} from the coefficient of $\bar\Psi_n$.

Then, we can define real modes (under the Majorana conjugate) for $n \ge 0$:
\begin{equation}
	\chi_n
		:= \frac{1}{\sqrt{2}} (\Psi_{n} + \Psi_{-n}),
	\qquad
	\phi_n
		:= - \frac{\I}{\sqrt{2}} (\Psi_{n} - \Psi_{-n}),
\end{equation}
such that
\begin{equation}
	\conj{\chi_n}
		= C \chi_n,
	\qquad
	\conj{\phi_n}
		= C \phi_n.
\end{equation}
These modes form two orthonormal sets:
\begin{equation}
	\bracket{\chi_m}{\chi_n}
		= \bracket{\phi_m}{\phi_n}
		= \delta_{mn}, \qquad
	\bracket{\chi_m}{\phi_n}
		= 0.
\end{equation}
It is then straightforward to check that these modes satisfy the equations:
\begin{equation}
	D \chi_n
		= \I \lambda_n \, \phi_n,
	\qquad
	D \phi_n
		= - \I \lambda_n \, \chi_n
\end{equation}
since
\begin{align*}
	D \chi_n
		= \frac{1}{\sqrt{2}} \,
			D (\Psi_n + \Psi_{-n})
		= \frac{\lambda_n}{\sqrt{2}} \, \gamma_* (\Psi_n - \Psi_{-n})
		= \I \lambda_n \, \phi_n.
\end{align*}
Squaring this equation gives:
\begin{equation}
	D^2 \chi_n
		= \Lambda_n \chi_n,
	\qquad
	D^2 \phi_n
		= \Lambda_n \phi_n,
	\qquad
	\Lambda_n
		:= \lambda_n^2.
\end{equation}
This also implies:
\begin{equation}
	D^2 \Psi_n
		= \Lambda_n \Psi_n.
\end{equation}
This means that the $(\chi_n, \phi_n)$ are eigenfunctions of the second-order (Laplace-type) kinetic operators, but not of the Dirac operator.
Note that it should be related to the decomposition of a Majorana spinor into a Weyl spinor and its complex conjugate~\cite[sec.~3.4]{Freedman:2012:Supergravity}.

The eigenvalues are indexed by $n \in \N$ and sorted by ascending order:
\begin{equation}
	0 < m^2 \le \Lambda_{0} \le \Lambda_{1} \le \cdots
\end{equation}
In particular, there is no zero-mode if $m^2 > 0$ (\Cref{sec:zero-modes}).

The Majorana field is expanded on the real modes as
\begin{equation}
	\Psi
		= \sum_{n \ge 0} (b_n \chi_n + c_n \phi_n)
\end{equation}
where $(b_n, c_n)$ are real Grassmann variables such that
\begin{equation}
	a_n
		= \frac{1}{\sqrt{2}} (b_n + \I c_n),
	\qquad
	\adj{a_n}
		= \frac{1}{\sqrt{2}} (b_n - \I c_n),
\end{equation}
and we have the relation
\begin{equation}
	\adj{a_n} a_n
		= \frac{\I}{2} ( b_{n} c_{n} - c_{n} b_{n} )
		= \I \, b_{n} c_{n}.
\end{equation}
Let us stress that $\adj{a_n}$ and $a_n$ anti-commute.
Note also that the Dirac conjugate is
\begin{equation}
	\bar\Psi
		= \sum_{n \ge 0} (b_n \chi_n^t + c_n \phi_n^t) C.
\end{equation}

\subsubsection{Evaluation of the functional integral}

We want to compute the generating functional with source $\eta$:
\begin{equation}
	Z[\eta]
		:= \int \dd \Psi
			\exp\big( - S[\Psi] + \I \bracket{\bar\eta}{\Psi} \big),
\end{equation}
where the action can be written in terms of the inner-product \eqref{eq:inner-product} as:
\begin{equation}
	S[\Psi]
		= \bra{\Psi} D \ket{\Psi}.
\end{equation}

The source is decomposed as
\begin{equation}
	\eta
		= \sum_{n \ge 0} (u_n \chi_n + v_n \phi_n )
		= \sum_{n \ge 0} \big( s_n \Psi_n + s_{-n} \Psi_{-n} \big),
	\qquad
	s_{-n}
		= \adj{s_n},
\end{equation}
where
\begin{equation}
	s_n
		= \frac{1}{\sqrt{2}} (u_n + \I v_n),
	\qquad
	\adj{s_n}
		= \frac{1}{\sqrt{2}} (u_n - \I v_n).
\end{equation}
We have the relation:
\begin{equation}
	\adj{s_n} s_n
		= \I u_n v_n.
\end{equation}

We can evaluate the inner-product which appears in the path integral,
\begin{equation}
	\bra{\Psi} D \ket{\Psi}
		= 2 \I \sum_{n} \lambda_{n} b_{n} c_{n}
		= 2 \sum_{n} \lambda_{n} \adj{a_n} a_n.
\end{equation}
and also:
\begin{equation}
	\bracket{\eta}{\Psi}
		= \sum_{n \ge 0} (u_n b_n + v_n c_n )
		= \sum_{n \ge 0} (s_n \adj{a_n} + \adj{s_n} a_n ).
\end{equation}

The functional integral reads:
\begin{subequations}
\begin{align}
	Z[\eta]
		&
		= \int \prod_{n \ge 0} \dd b_n \dd c_n
			\exp \left(- \I \sum_{n \ge 0} \big[
				\lambda_n b_{n} c_{n}
				+ u_n b_n + v_n c_n
				\big]
				\right)
		\\ &
		= \int \prod_{n \ge 0} \dd a_n \dd \adj{a_n}
			\exp \left(
				\sum_{n \ge 0} \big[
					- \lambda_n \adj{a_n} a_n
					+ \I \, s_n \adj{a_n} + \I \, \adj{s_n} a_n
				\big]
				\right).
\end{align}
\end{subequations}
The next step consists in shifting the variables $a_n$:
\begin{equation}
	\bar a_n
		= a_n + \frac{\I}{\lambda_n} \, s_n,
	\qquad
	\adj{\bar a_n}
		= \adj{a_n} - \frac{\I}{\lambda_n} \, \adj{s_n}
\end{equation}
such that
\begin{equation}
	Z[\eta]
		= \exp \left( \sum_{n \ge 0}
				\frac{1}{\lambda_n} \, \adj{s_n} s_n
				\right)
			\int \prod_{n \ge 0} \dd \bar a_n \dd \adj{\bar a_n} \exp \left(
				- \sum_{n \ge 0} \lambda_n \adj{\bar a_n} \bar a_n
				\right).
\end{equation}
The integral is a simple Gaussian integral of complex Grassmann variables:
\begin{equation}
	\label{eq:Z-source-modes}
	Z[\eta]
		= \exp \left( \sum_{n \ge 0}
				\frac{1}{\lambda_n} \, \adj{s_n} s_n
				\right)
			\prod_{n \ge 0} \lambda_n.
\end{equation}
Note that only half of the eigenvalues are included because we had to combine the real functions into complex functions, which lifts the double degeneracy that one has with a Dirac fermion.
The product of the positive eigenvalues gives the square-root of the determinant:
\begin{equation}
	\label{eq:infinite-det}
	\prod_{n \ge 0} \lambda_n
		= \sqrt{\prod_{n \ge 0} \lambda_n^2}
		= \bigg( \prod_{n \in \Z} \lambda_n^2 \bigg)^{\frac{1}{4}}
		= \big( \det D^2 \big)^{1/4}
		= \det \left( - \Delta + \frac{R}{4} + m^2 \right)^{1/4}.
\end{equation}
The first equality allows writing squares of eigenvalues, such that one can extend the range to negative $n$ after the second equality since $\lambda_{-n} = - \lambda_n$.
Using formal manipulations of determinants, we can rewrite $\sqrt{\det D^2} = \det D$ such that:
\begin{equation}
	\prod_{n \ge 0} \lambda_n
		= \sqrt{\det D}
		= \sqrt{\det \left( \I \slashed\grad + m \gamma_* \right)}.
\end{equation}
Note that the fact that one can take the square-root without ambiguity (up to a sign) is a consequence of the self-adjointness of the operator~\cite[p.~1470]{Blau:1989:DeterminantsDiracOperators}.

The Green function corresponds to
\begin{equation}
	\begin{aligned}
	S(x, y)
		&
		:= \bra{x} \frac{1}{D} \ket{y}
		:= \bra{x} \frac{1}{\I \slashed\grad + m \gamma_*} \ket{y}
		\\ &
		= \sum_{n \in \Z} \frac{1}{\lambda_n} \,
			\Psi_n(x) \adj{\Psi_n(y)}
		= \I \sum_{n \ge 0} \frac{1}{\lambda_n} \,
			\big( \phi(x) \chi(y)^t - \chi(x) \phi(y)^t \big).
	\end{aligned}
\end{equation}
It follows from:
\begin{equation}
	\bra{x} \frac{1}{D} \ket{y}
		= \sum_{n \in \Z} \bra{x}
			\frac{1}{D} \ket{\Psi_n}
			\bracket{\Psi_n}{y}
		= \sum_{n \in \Z} \frac{1}{\lambda_n} \,
			\bracket{x}{\Psi_n}
			\bracket{\Psi_n}{y}.
\end{equation}
The Green function is antisymmetric and purely imaginary:
\begin{equation}
	S_{\alpha\beta}(x, y)
		= - S_{\beta\alpha}(y, x).
\end{equation}
In full similarity, we obtain the Green function of $D^2$:
\begin{equation}
	\label{eq:green-func-deriv}
	G(x, y)
		:= \bra{x} \frac{1}{D^2} \ket{y}
		= \sum_{n \in \Z} \frac{1}{\Lambda_n}
			\Psi_n(x) \adj{\Psi_n(y)}.
\end{equation}
Note that $S$ and $G$ are $2$-dimensional matrices in terms of Dirac indices.
The trace over Dirac indices is denoted by $\tr_D$.

Finally, we have that
\begin{equation}
	\bra{\eta} S \ket{\eta}
		= \sum_{n \in \Z} \frac{1}{\lambda_n} \,
			\adj{s_n} s_n
		= 2 \I \sum_{n \ge 0} \frac{1}{\lambda_n} \,
			u_n v_n.
\end{equation}

As a conclusion, we find:
\begin{equation}
	\label{eq:Z-source-op}
	Z[\eta]
		= \exp \left(\frac{1}{2} \bra{\eta} S \ket{\eta} \right)
			\det \left( - \Delta + \frac{R}{4} + m^2 \right)^{1/4}.
\end{equation}
The factor of $1/2$ arises because the Green function has a sum $n \in \Z$, but the functional integral gives only $n \ge 0$ in the exponential.

This gives the expression of the effective action
\begin{equation}
	S_{\text{eff}}
		:= - \ln Z[0]
		= - \frac{1}{4} \, \ln \det D^2
		= - \frac{1}{4} \, \tr \ln D^2
		= - \frac{1}{4} \, \tr \ln \left(-\Delta + \frac{R}{4} + m^2 \right).
\end{equation}
This determinant can be defined using the standard heat kernel and zeta function methods, since $D^2$ is an operator of Laplace type.

Let us pause to comment on the case where there are zero-modes (vanishing eigenvalues, or massless fermion).
In this case, $Z[\eta]$ in \eqref{eq:Z-source-modes} looks ill-defined because the first term diverges and the product of eigenvalues vanishes.
However, the eigenvalues $\lambda_n = 0$ do not appear in the sum in $\mean{\Psi | D \Psi}$ such that the product would be only over strictly positive eigenvalues, $n > 0$.
Similarly, \eqref{eq:Z-source-op} has instead $\det' D$ and $\widetilde S$, the determinant and Green functions without zero-modes.
Moreover, the integrals over zero-modes would remain to be done: since they are fermionic, it looks like the result would vanish.
This is solved by inserting zero-modes in the functional integral and carefully normalizing~\cite{Blau:1989:DeterminantsDiracOperators, Erbin:2019:IntroductionStringField}.

\section{Spectral analysis}
\label{sec:spectral-analysis}

In this section, we review some points of spectral analysis that will be used to compute the gravitational action.
In \Cref{sec:spectral-functions}, we define the spectral functions (Green's function, heat kernel and zeta functions) and recall some of their properties.
Then, \Cref{sec:spectral:conformal} describes how the objects are affected by a Weyl transformations.
Finally, in \Cref{sec:zero-modes}, we discuss the zero-modes of the Dirac operator.

While it is possible to derive most formulas for both Green functions $S(x,y)$ and $G(x,y)$, we will not need any formula for $S(x,y)$ in our approach, and hence we focus on the Green function $G(x, y)$ of the operator $D^2$.

We define the area of the surface with metric $g$ as
\begin{equation}
	A[g]
		= \int \dd^2 x \, \sqrt{g}.
\end{equation}

\subsection{Spectral functions}
\label{sec:spectral-functions}

From now on, we will denote collectively the spinors $\chi_n$ and $\phi_n$ by $\Psi_n$, since they are associated to the same eigenvalue $\Lambda_n = \lambda_n^2$ of the operator $D^2 = - \Delta + R/4 + m^2$.
States associated with the eigenvalue $\Lambda_0 = m^2$ are called zero-modes.
Since both $\lambda_0 = \pm m$ lead to the same $\Lambda_0$, we denote by $N_0$ the number of such pairs, and write $\Psi_{0,i}$ the eigenstates with $\lambda_0 = + m$.
We can also understand $N_0$ as counting the number of complex spinors (under Majorana conjugation), given \eqref{eq:mode-Psi-n-neg}, or the number of eigenmodes with positive (or negative) chirality~\cite{DHoker:1988:GeometryStringPerturbation}.
Indeed, $\com{(\I \slashed\grad)^2}{\gamma_*} = 0$ such that it is possible to use a basis of zero-modes with definite chirality $\gamma_* \Psi_{0,i,\pm} = \pm \Psi_{0,i,\pm}$.

We will put a tilde on every quantity from which the contributions of the zero-modes have been subtracted, and a subscript “$(0)$” when $m=0$.
In particular, $\Lambda_n$ is related to $\Lambda_n^{(0)}$ by
\begin{equation}
	\label{eq:eigenvalues-massive-massless}
	\Lambda_n = \Lambda_n^{(0)} + m^2.
\end{equation}
This shows that there is no zero-mode when $m^2 \neq 0$.

\paragraph{Green's function}

The Green's function for the operator $D^2$ is given by \eqref{eq:green-func-deriv}
\begin{equation}
	\label{def:Green-function-fermions}
	D^2 G(x, y) = \frac{\delta(x-y)}{\sqrt g} \, \mathrm I_2
	\quad \Leftrightarrow \quad
	G(x, y) = \sum_{n \in \Z} \frac{\Psi_n(x) \Psi_n(y)^\dagger}{\Lambda_n},
\end{equation}
where $\mathrm I_2$ denotes the $2$-dimensional identity.
$G$ and its hermitian conjugate $G^\dagger$ are related by
\begin{equation}
 	G(y, x) = G(x, y)^\dagger.
\end{equation}
One can define the Green's function without the zero-modes by
\begin{equation}
	\label{eq:G-Gtilde}
	\tilde G(x, y) = \sum_{n \neq 0} \frac{\Psi_n(x) \Psi_n(y)^\dagger}{\Lambda_n} = G(x, y) - \frac{1}{m^2} \, P(x, y),
\end{equation}
where
\begin{equation}
	\label{eq:proj-zero-mode-orthonormal}
	P(x, y) = \sum_i \Psi_{0, i}(x) \Psi^\dagger_{0, i}(y)
\end{equation}
is the projector on the zero-mode subspaces, with $\{ \psi_i(x) \}$ an orthonormal basis (see \Cref{sec:spectral:conformal} for a discussion of this point).
It satisfies the following equation
\begin{equation}
	D^2 \tilde G(x, y) = \frac{\delta(x-y)}{\sqrt g} \, \mathrm I_2 - P(x, y),
\end{equation}
and it is orthogonal to the projector
\begin{equation}
	\int \dd^2 z \sqrt{g} \, P(x, z) \tilde G(z, y) = 0.
\end{equation}
The last relation also implies
\begin{equation}
	\tilde G(x, y)
		= \int \dd^2 z \sqrt{g} \, \tilde G(x, z) D_z^2 \tilde G(z, y)
		= \int \dd^2 z \sqrt{g} \, D_z \tilde G(x, z) D_z \tilde G(z, y).
\end{equation}

When $x$ goes to $y$, the Green's function presents the usual logarithmic singularity
\begin{equation}
	\label{eq:Gsing}
	G_{\text{sing}} = \tilde G_{\text{sing}} = -\frac{1}{4\pi} \ln(\ell(x, y)^2)
\end{equation}
where $\ell(x, y)$ is the geodesic distance between $x$ and $y$.
A regularization of the Green's function at coincident points is then
\begin{equation}
	G_R(x) = \lim_{y \to x} \left(G(x, y) + \frac{1}{4\pi} \ln \left(\mu^2 \ell^2(x, y)\right)\right),
\end{equation}
where $\mu$ is a mass scale needed for dimensional reasons.

The Green function $G$ is ill-defined when $\Lambda_0 = m = 0$  since it contains a sum over all eigenmodes with coefficients containing $\Lambda_n^{-1}$.

\paragraph{Heat kernel}

The heat kernel is defined to be the solution of
\begin{equation}
	\label{eq:def-heat-kernel}
	\left(\frac{\dd}{\dd t} + D^2\right) K(t, x, y) = 0, \qquad K(t, x, y) \underset{t\to 0}{\sim} \frac{\delta(x-y)}{\sqrt g} \, \mathrm I_2.
\end{equation}
In terms of eigenvectors and eigenvalues, this can be expressed as
\begin{equation}
	\label{eq:heat-kernel-eigenvectors}
	K(t, x, y) = \sum_{n \in Z} \e^{-\Lambda_n t} \Psi_n (x) \Psi_n(y)^\dagger.
\end{equation}
One also defines the integrated heat kernel by
\begin{equation}
	\label{eq:integrated-heatkernel}
	K(t) = \int \dd^2 x \sqrt g \, \tr_D K(t, x, x) = \sum_{n \in \Z} \e^{-\Lambda_n t}.
\end{equation}
The corresponding quantities $\tilde K(t)$ and $\tilde K(t, x, y)$ are obtained by excluding the zero-modes from the sum.
In particular, if $m = 0$, one needs to work with these quantities.

For $t>0$, we see from \eqref{eq:heat-kernel-eigenvectors} that $K(t,x,y)$ is given by a converging sum and is finite, even as $x\to y$.
For $t\to 0$ various divergences are recovered, in particular,
\begin{equation}
	\label{KG}
	\int_0^\infty \dd t \, K(t,x,y) = G(x,y)
\end{equation}
exhibits the short-distance singularity of the Green's function \eqref{eq:Gsing}.

The behavior of $K$ for small $t$ is related to the asymptotics of the eigenvalues $\Lambda_n$ and eigenfunctions $\Psi_n$ for large $n$, which in turn is related to the short-distance properties of the Riemann surface.
It is well-known that the small-$t$ asymptotics is given in terms of local expressions of the curvature and its derivatives, and that on a compact manifold without boundaries one has
\begin{equation}
	\label{Kasymp}
	K(t,x,y) \underset{t\to 0}{\sim} \frac{1}{4\pi t} \, \e^{-\ell^2(x,y)/4t } \sum_{k\geq 0} t^k a_k(x,y).
\end{equation}
The expansion coefficients can be computed recursively using normal coordinates around $x$ (since for small $t$, the exponential forces $\ell^2$ to be small).
To compute the gravitational action, we will only need the first ones at coinciding points, which for our operator $D^2$ turn out to be
\begin{subequations}
	\label{heat-kernel-coefficients-spinor}
	\begin{align}
		a_0(x, x) &= \mathrm I_2 \\
		a_1(x, x) &= \left(-\frac{R}{12} - m^2\right) \mathrm I_2 \label{heat-kernel-a1}.
	\end{align}
\end{subequations}
For small $t$, $K$ is then given by
\begin{equation}
	\label{K-t-small}
	K(t) = \frac{A}{2\pi t} - \frac{1}{24\pi} \int \dd^2 x \sqrt g \, R(x) - \frac{m^2 A}{2\pi} + o(t).
\end{equation}

Another useful relation is
\begin{equation}
	\label{eq:relation-K-K0}
	\tilde K(t) = \e^{- m^2 t} \tilde K^{(0)}(t)
\end{equation}
using \eqref{eq:eigenvalues-massive-massless}.

\paragraph{Zeta functions}

The zeta function and its integrated version are defined by
\begin{equation}
	\label{eq:zeta}
	\zeta(s, x, y)
		= \sum_n \frac{\Psi_n(x) \Psi_n(y)^\dagger}{\Lambda_n^s},
	\qquad
	\zeta(s)
		= \int \dd^2 x \, \tr_D \zeta(s, x, x)
		= \sum_n \frac{1}{\Lambda_n^s}.
\end{equation}
They are related to the heat kernel by a Laplace transform
\begin{equation}
	\label{eq:zeta-heat-kernel}
	\zeta(s, x, y) = \frac{1}{\Gamma(s)} \int_0^{+\infty} \dd t \, t^{s-1} K(t, x, y),
	\qquad
	\zeta(s) = \frac{1}{\Gamma(s)} \int_0^{+\infty} \dd t \, t^{s-1} K(t).
\end{equation}
Plugging \eqref{K-t-small} into \eqref{eq:zeta-heat-kernel} enables one to compute $\zeta(0)$:
\begin{equation}
	\zeta(0)
		= - \frac{1}{24\pi} \int \dd^2 x \sqrt g \, R(x) - \frac{m^2 A}{2\pi}
		= - \frac{1}{24\pi} \int \dd^2 x \sqrt g \, R(x) - \frac{m^2 A}{2\pi}
\end{equation}
In the same way, one shows that the zeta function has a pole for $s=1$ with residue $\frac{a_0(x, x)}{4\pi} = \frac{\mathrm I_2}{4\pi}$.
Then, using the relation
\begin{equation}
	\tilde\zeta(0, x, x)
		= \zeta(0, x, x) - P(x, x),
\end{equation}
we obtain
\begin{equation}
	\tilde\zeta(0) = \zeta(0) - N_0.
\end{equation}

The zeta function can be used to define a regularized version of the Green's function at coincident points:
\begin{equation}
	\label{eq:G-zeta}
	G_\zeta(x) = \lim_{s \to 1} \left(\mu^{2s-2} \zeta_{\text{reg}}(s, x, x) \right),
	\qquad
	\zeta_{\text{reg}}(s, x, x) = \zeta(s, x, x) - \frac{\mu^{2-2s} \, \mathrm I_2}{4\pi(s-1)},
\end{equation}
and the same for $\tilde \zeta$ and $\tilde G_\zeta$ where $G_\zeta$ and $\tilde G_\zeta$ are related by
\begin{equation}
	G_\zeta(x) = \tilde G_\zeta(x) + \frac{1}{m^2} \, P(x, x).
\end{equation}
$G_R$ and $G_\zeta$ only differ by a constant~\cite{Bilal:2013:MultiLoopZetaFunction}:
\begin{equation}
	G_\zeta(x) = G_R(x) + \alpha.
\end{equation}

Finally, we will use the zeta function to provide a regularized definition of the logarithm of the determinant of the operator $D^2$.
Let consider
\begin{equation}
	S_{\text{eff}}[g] = - \ln Z_m[g] = - \frac{1}{4} \sum_{n \neq 0} \ln \frac{\Lambda_n}{\hat\mu^2},
\end{equation}
where the energy scale $\hat \mu$ has been introduced to ensure that the argument of the logarithm is dimensionless.
The infinite sum is regularized through an analytic continuation such that:
\begin{equation}
	\sum_{n \neq 0} \ln \frac{\lambda_n}{\mu^2} = - \zeta'(0) - \ln \hat\mu^2 \zeta(0),
\end{equation}
and then
\begin{equation}
	\label{zeta-function-regularization}
	S_{\text{eff}} = \frac{1}{4} \zeta'(0) + \frac{1}{4} \ln \hat\mu^2 \zeta(0).
\end{equation}

\subsection{Weyl transformations}
\label{sec:spectral:conformal}

A Weyl transformation corresponds to a rescaling of the metric
\begin{equation}
	g \longrightarrow \e^{2\omega} g.
\end{equation}
It can be used to reach the conformal gauge
\begin{equation}
	g = \e^{2\sigma(x)} \hat g,
\end{equation}
where $\hat g$ is a fixed metric and $\sigma$ is the Liouville field.
If the latter is small, one finds
\begin{equation}
	\delta g = 2 \delta \sigma(x) \hat g.
\end{equation}

The metric in which the Green functions, heat kernel and zeta functions are defined is indicated by an index, for example $K_g(t)$ and $K_{\hat g}(t)$.
In the case of a functional, it is given in a bracket.
Finally, differential operators receive a hat, for example $\hat\grad_\mu$ is expressed in the metric $\hat g$.

Using the variation of \eqref{def:Green-function-fermions} under an infinitesimal Weyl transformation and the expression for $\delta D^2$ given in \eqref{eq:variation-D2}, we see that, at first order in $\delta \sigma$, $\delta G$ is a solution of
\begin{equation}
	D^2 \delta G(x, y) = -2 m^2 \delta \sigma(x) G(x, y) + \partial_\nu \delta \sigma(x) \gamma^{\mu\nu} \nabla_\mu G(x, y) + \frac{1}{2} \Delta(\delta \sigma(x)) G(x, y).
\end{equation}
The solution is obtained by convolution with $G$:
\begin{equation}
	\label{eq:deltaG}
	\begin{aligned}
		\delta G(x, y)
			= &- 2 m^2 \int \dd^2 z \sqrt g \, G(x, z) \delta \sigma(z) G(z, y) + \int \dd^2 z \sqrt g \, G(x, z) \partial_\nu \delta \sigma(z) \gamma^{\mu\nu} \nabla_\mu G(z, y)
			\\
			&+ \frac{1}{2} \int \dd^2 z \sqrt g \, G(x, z) \Delta(\delta \sigma(z)) G(z, y).
	\end{aligned}
\end{equation}
We can simplify this expression as follows (writing $G(x, y)$ as $G_{xy}$ for concision):
\begin{align*}
	\delta G_{xy}
		&
		= - \int \dd^2 z \sqrt{g} \, G_{xz} \left(
			2 m^2 \, \delta\sigma_z
			- (\pd_z^\mu \delta\sigma_z) \grad_{z\mu}
			+ (\slashed\pd_z \delta\sigma_z) \slashed\grad_z
			- \frac{1}{2} \Delta_z \delta\sigma_z
			\right) G_{zy}
		\\ &
		\begin{aligned}
		= - \int \dd^2 z \sqrt{g} \, \delta\sigma_z \bigg[
			&
			2 m^2 \, G_{xz} G_{zy}
			+ \cancel{G_{xz} \overleftarrow{\grad}_{z\mu} \grad_{z\mu} G_{zy}}
			+ G_{xz} \Delta_z G_{zy}
			\\ &
			- G_{xz} \overleftarrow{\slashed\grad}_z \slashed\grad_z G_{zy}
			- G_{xz} \slashed\grad_z^2 G_{zy}
			- \frac{1}{2} \, G_{xz} \overleftarrow{\Delta}_z G_{zy}
			\\ &
			- \cancel{G_{xz} \overleftarrow{\grad}_{z \mu} \grad_{z}^{\mu} G_{zy}}
			- \frac{1}{2} \, G_{xz} \Delta_z G_{zy}
			\bigg]
		\end{aligned}
		\\ &
		\begin{aligned}
		= - \int \dd^2 z \sqrt{g} \, \delta\sigma_z \bigg[
			&
			m^2 \, G_{xz} G_{zy}
			+ \frac{1}{2} \, G_{xz} D_z^2 G_{zy}
			+ \frac{1}{2} \, G_{xz} \overleftarrow{D}^2_z G_{zy}
			- G_{xz} \overleftarrow{\slashed\grad}_z \slashed\grad_z G_{zy}
			\bigg]
		\end{aligned}
		\\ &
		\begin{aligned}
		= - \int \dd^2 z \sqrt{g} \, \delta\sigma_z \bigg[
			&
			m^2 \, G_{xz} G_{zy}
			+ \frac{1}{2} \, G_{xz} \frac{\delta_{zy}}{\sqrt{g}}
			+ \frac{1}{2} \, \frac{\delta_{xz}}{\sqrt{g}} G_{zy}
			- G_{xz} \overleftarrow{\slashed\grad}_z \slashed\grad_z G_{zy}
			\bigg],
		\end{aligned}
\end{align*}
where we used $\gamma_{\mu\nu} = \gamma_\mu \gamma_\nu - g_{\mu\nu}$ to get the first equality, integrated by part the derivatives acting on $\delta \sigma_z$ for the second equality, and wrote $\Delta$ and $\slashed\grad^2$ in terms of $D^2$ for the third equality.
We defined $\overleftarrow{\slashed\grad}_z$ and other similar symbols such that its derivatives act on the object closest on its left, with gamma matrices staying where the symbol sits.
Finally, we can use the Green equation \eqref{def:Green-function-fermions} to get:
\begin{equation}
	\begin{aligned}
	\delta G(x, y)
		= &- \frac{1}{2} \big(\delta\sigma(x) + \delta\sigma(y)\big) G(x, y)
			- m^2 \int \dd^2 z \sqrt{g} \, \delta\sigma(z) \, G(x, z) G(z, y)
			\\
			&+ \int \dd^2 z \sqrt{g} \, \delta\sigma(z) \, G(x, z) \overleftarrow{\slashed\grad}_z \slashed\grad_z G(z, y).
	\end{aligned}
\end{equation}
Using the fact that
\begin{equation}
	\begin{gathered}
	\delta \ell^2(x, y) = \ell^2(x, y) (\delta \sigma(x) + \delta \sigma(y)) + \mc O(\ell^4(x, y))
	\\
	\Longrightarrow \quad
	\lim_{y\to x} \delta \ln (\mu^2\ell^2(x, y)) = 2 \delta \sigma(x),
	\end{gathered}
\end{equation}
one can compute the variation of the regularized Green functions
\begin{equation}
	\label{eq:var-G-zeta}
	\delta G_\zeta(x) = \delta G_R(x) = \delta \lim_{y \to x} \left( G(x, y) + \frac{1}{4\pi} \ln \left(\mu^2 \ell^2(x, y)\right)\right).
\end{equation}

The zero-modes transform under a Weyl transformation.
Given the transformation law \eqref{eq:transformation-slashed-nabla} of the Dirac operator, it follows directly that if $\Psi_{0}$ is a zero-mode of $\slashed\grad$, then
\begin{equation}
	\label{eq:var-zero-mode}
	\hat\Psi_{0}
		= \e^{\sigma/2} \Psi_{0}
\end{equation}
is a zero-mode of $\hat{\slashed\grad}$.
A first consequence is that a Weyl transformation does not change the number of zero-modes $N_0$.

This, together with the fact that the integration is over a different space, implies that the zero-modes in the $\hat g$ metric will not be orthonormal even if the zero-modes in the metric $g$ formed an orthonormal basis~\cite{Blau:1989:DeterminantsDiracOperators}.\footnotemark{}
\footnotetext{%
	\label{ft:scalar-zero-mode}
	The scalar field $\phi$ provides a simple example.
	Since the scalar Laplacian is Weyl invariant in two dimensions, the zero-mode is also invariant and $\phi_0 = \hat\phi_0$.
	However, if $\phi_0 = 1/\sqrt{A}$ is normalized in the first metric, it is not in the second metric and $\kappa[\hat g] = \int \dd^2 x \sqrt{\hat g} \, \phi_0^2 = \hat A / A$.
	It would have been possible to start with the non-normalized zero-mode $\phi_0 = 1$, such that $\kappa[g] = A$ and $\kappa[\hat g] = \hat A$.
	This motivates the introduction of the normalization in the projection operator: its absence in~\cite{Bilal:2017:2DQuantumGravity} was not important because there was a single zero-mode which expression is known, and thus the variation in the normalization could be tracked by hand.
	The situation is much more complicated for a Majorana field.
}%
As a consequence, the projector \eqref{eq:proj-zero-mode-orthonormal} on the zero-modes must be modified.
For this, reason, we introduce the normalization matrix
\begin{equation}
	\label{eq:normalization-matrix}
	\kappa_{ij}[g] = \int \dd^2 x \sqrt{g} \, \tr_D \Psi_{0,i}(x)^\dagger \Psi_{0,j}(x).
\end{equation}
Then the projector onto the zero-modes becomes
\begin{equation}
	\label{eq:proj-zero-mode}
	P_g(x, y) = \sum_{i,j} \Psi_{0, i}(x) \kappa^{ij}[g] \Psi^\dagger_{0, j}(y),
\end{equation}
where $\kappa^{ij}$ denotes the inverse of $\kappa_{ij}$.
The trace of the projector counts the number of zero-modes
\begin{equation}
	\label{eq:trace-projector}
	\int \dd^2 x \sqrt{g} \, \tr_D P_g(x, x)
		= N_0
\end{equation}
since
\begin{equation}
	\int \dd^2 x \sqrt{g} \, \tr_D P_g(x, x)
		= \sum_{i,j} \int \dd^2 x \sqrt{g} \, \tr_D \Psi_{0, i}(x) \kappa^{ij}[g] \Psi^\dagger_{0, j}(y)
		= \kappa^{ij}[g] \kappa_{ij}[g]
		= \delta^i_i.
\end{equation}
The following relation will be useful:
\begin{equation}
	\label{eq:variation-kappa}
	\delta \ln \det \kappa = \int \dd^2 x \sqrt{g} \, \delta \sigma(x) \tr_D P_g(x, x),
\end{equation}
which follows from
\begin{equation}
	\delta \tr \ln \kappa
		= \tr \kappa^{-1} \delta \kappa
		= \kappa^{ij} \int \dd^2 x \sqrt{g} \, \delta \sigma \tr_D \Psi_{0,i}(x) \Psi_{0,j}(x)^\dagger.
\end{equation}

\subsection{Spin structure and zero-modes of the Dirac operator}
\label{sec:zero-modes}

In this section, we discuss the zero-modes of the massless operator $- \Delta + \frac{R}{4}$, which are also the zero-modes of $\slashed\nabla$, in terms of the spin structure.

\paragraph{Sphere}

On the sphere, we can take for $g$ the round metric, which has $R > 0$ constant.
As $-\Delta$ is a positive operator, we see that $- \Delta + \frac{R}{4}$ cannot have any zero-mode and hence
\begin{equation}
	N_0 = 0,
	\qquad
	P_g(x, y)
		= 0.
\end{equation}

\paragraph{Torus}

On the torus, we can use the flat metric $g = \delta$.
Then, a zero-mode satisfies
\begin{equation}
	\slashed\pd \Psi_0 = 0
\end{equation}
and the only solutions are the two constant spinors
\begin{equation}
	\label{eq:torus-zero-modes}
	\Psi_0
		= \frac{1}{\sqrt A}
		\begin{pmatrix}
			1 \\
			0
		\end{pmatrix},
		\qquad
	\Psi_{0}'
		= \frac{1}{\sqrt A}
		\begin{pmatrix}
			0 \\
			1
		\end{pmatrix},
\end{equation}
where $A := A[\delta]$ is the area of the torus.
The solution survives only for the odd spin structure, which has periodic boundary conditions in both directions.
If at least one boundary has anti-periodic conditions, then there is no solution.

As a consequence, there is one (complex) zero-mode for the odd spin structure, and no zero-mode for the three even spin structures:
\begin{equation}
	N_0
		= 0
		\quad (\text{even structures}),
	\qquad
	N_0
		= 1
		\quad (\text{odd structure}).
\end{equation}

We immediately derive the normalization matrix and projector:
\begin{equation}
	\label{eq:torus-kappa-proj}
	P_g(x, y)
		= \frac{1}{A}
		\begin{pmatrix}
			1 & 0 \\
			0 & 1
		\end{pmatrix},
	\qquad
	\kappa[g]
		=
			\begin{pmatrix}
				1 & 0 \\
				0 & 1
			\end{pmatrix},
\end{equation}
such that
\begin{equation}
	- \frac{1}{2} \, \ln \det \frac{\kappa[g]}{\kappa[\hat g]}
		= \ln \frac{A}{\hat A}.
\end{equation}

\paragraph{Higher-genus}

For a genus $g \ge 2$, there is generically one (resp.\ no) zero-mode when the spin structure is odd (resp.\ even).
However, there can be up to $g$ zero-modes~\cite{DHoker:1988:GeometryStringPerturbation}.
Zero-modes can be computed by taking $\hat g$ to be the metric such that $\hat R = - 1$ or to be the canonical metric.

\section{Gravitational action}
\label{sec:gravitational-action}

Since it is complicated to evaluate the effective action \eqref{zeta-function-regularization} directly, it is simpler to consider the gravitational action \eqref{eq:def-gravitationnal-action}
\begin{equation}
	S_{\text{grav}}[\hat g, g] = \frac{1}{4}(\zeta_g'(0) + \ln \mu^2 \zeta_g(0)) - \frac{1}{4}(\zeta_{\hat g}'(0) + \ln \mu^2 \zeta_{\hat g}(0)).
\end{equation}
To simplify further the computation, one can consider an infinitesimal difference such that
\begin{equation}
	\delta S_{\text{grav}}[g] = \frac{1}{4}(\delta \zeta_g'(0) + \ln \mu^2 \delta \zeta_g(0)).
\end{equation}

Finally, it is natural to consider both metrics to be related by a Weyl transformation $g = \e^{2\sigma(x)} \hat g$, or its infinitesimal version $g = 2 \delta \sigma(x) \hat g$.
The gravitational action will be recovered at the end by integrating over $\delta\sigma$.

\subsection{Variation of the zeta function}

To compute the gravitational action, we need $\delta \zeta(0)$ and $\delta\zeta'(0)$.
The variation of the zeta function is given by
\begin{equation}
 \delta\zeta(s) = -s \sum_n \frac{\delta\Lambda_n}{\Lambda_n^{s+1}}.
\end{equation}
As there is no zero-mode contribution to this equation, we can work with the function $\tilde \zeta$ instead of $\zeta$:
\begin{equation}
	\delta \zeta(s) = \delta \tilde \zeta(s).
\end{equation}
Doing this, we do not have to distinguish between the massive and the massless case.

The variation of the non-zero eigenvalues $\Lambda_n$ can be obtained through usual perturbation theory:
\begin{equation}
	\begin{aligned}
		\delta \Lambda_n
			&= \langle\Psi_n | \delta D^2 | \Psi_n \rangle
			\\
			&= -2 (\Lambda_n - m^2) \langle \Psi_n | \delta\sigma | \Psi_n \rangle - \langle \Psi_n | \partial_\nu (\delta \sigma) \gamma^{\mu\nu} \nabla_\mu | \Psi_n \rangle - \frac{1}{2}\langle \Psi_n | \Delta (\delta \sigma) | \Psi_n \rangle,
	\end{aligned}
\end{equation}
where we used \eqref{eq:variation-D2}.
However, the first eigenvalue $\Lambda_0 = m^2$ does not change: $\delta \Lambda_0 = 0$.
We then get
\begin{align}
	\label{eq:variation-zeta}
	\delta\tilde\zeta(s)
		&
		= 2s \sum_{n\neq 0} \int \dd^2 x \sqrt g \, \delta\sigma(x) \frac{\Psi_n(x)^\dagger \Psi_n(x)}{\Lambda_n^s}
		\nonumber
		\\ & \qquad
		- 2 m^2 s \sum_{n \neq 0} \int \dd^2 x \sqrt g \, \delta\sigma(x) \frac{\Psi_n(x)^\dagger \Psi_n(x)}{\Lambda_n^{s+1}}
		\nonumber
		\\ & \qquad
		+ s \sum_{n \neq 0} \int \dd^2 x \sqrt g \, \frac{\partial_\nu (\delta \sigma) \Psi_n(x)^\dagger \gamma^{\mu\nu} \nabla_\mu \Psi_n(x)}{\Lambda_n^{s+1}}
		\nonumber
		\\ & \qquad
		+ \frac{s}{2} \sum_{n\neq 0} \int \dd^2 x \sqrt g \, \Delta (\delta \sigma) \frac{\Psi_n(x)^\dagger \Psi_n(x)}{\Lambda_n^{s+1}}
	\\ &
	=: 2s \, I_1 - 2m^2 s \, I_2 + s \, I_3 + \frac{s}{2} \, I_4.
\end{align}

One can show that
\begin{subequations}
	\label{eq:I-values}
	\begin{align}
		&
		I_1
			= \int \dd^2 x \sqrt g \, \delta\sigma(x) \tr_D \tilde \zeta(s, x, x),
		\\ &
		I_2
			= \int \dd^2 x \sqrt g \, \delta\sigma(x) \tr_D \tilde \zeta(s+1, x, x),
		\\ &
		I_3
			= - \frac{1}{2} \int \dd^2 x \sqrt g \, \delta \sigma(x) \tr_D (\Delta \tilde \zeta(s+ 1, x, x)),
		\\ &
		I_4
			= \int \dd^2 x \sqrt g \, \delta \sigma \tr_D(\Delta \tilde \zeta(s+1, x, x)).
	\end{align}
\end{subequations}
We define the Laplacian acting on the zeta function below, see \eqref{eq:laplacian-zeta}.
The first two expressions are immediate consequences of the definition \eqref{eq:zeta} of the zeta function.
Moreover, $I_4$ follows directly from an integration by parts, and only $I_3$ is non-trivial.

The first step is to rewrite the derivative acting on $\delta \sigma$ as a covariant derivative, and then integrate by parts:
\begin{equation}
	\label{eq:third-term-variation-zeta}
	\begin{aligned}
	I_3
		&
		= \sum_{n \neq 0} \frac{1}{\Lambda_n^{s+1}} \int \dd^2 x \sqrt g \, \delta \sigma
			\Bigl(
				\Psi_n(x)^\dagger \overleftarrow{\nabla}_\nu \gamma^{\mu\nu} \nabla_\mu \Psi_n(x)
				+ \Psi_n(x)^\dagger \gamma^{\mu\nu} \nabla_\nu \nabla_\mu \Psi_n(x)
			\Bigr)
		\\ &
		= \sum_{n \neq 0} \frac{1}{\Lambda_n^{s+1}} \int \dd^2 x \sqrt g \, \delta \sigma
		\Bigl(
			\Psi_n(x)^\dagger \overleftarrow{\slashed\nabla} \slashed\nabla \Psi_n(x)
			+ \Psi_n(x)^\dagger \overleftarrow{\nabla}^\mu \nabla_\mu \Psi_n(x)
			\\ & \hspace{4.5cm}
			- \frac{R}{4} \, \Psi_n(x)^\dagger \Psi_n(x)
		\Bigr).
	\end{aligned}
\end{equation}
The second equality follows using $\gamma^{\mu\nu} = \gamma^\mu \gamma^\nu - g^{\mu\nu}$ and \eqref{eq:antisym-deriv}.
We will now compute the first term, before turning our attention to the second and third terms together.

We notice that, for $n \neq 0$, $\slashed\nabla \Psi_n$ is also an eigenvector of $D^2$ associated with the eigenvalue $\Lambda_n$ and normalized such that
\begin{equation}
	\int \dd^2 x \sqrt g \, (\slashed \nabla \Psi_n)^\dagger \slashed \nabla \Psi_n
		= - \int \dd^2 x \sqrt g \, \Psi_n^\dagger \slashed\nabla^2 \Psi_n
		= \Lambda_n^{(0)}
		= \Lambda_n - m^2.
\end{equation}
As the heat kernel is uniquely defined by \eqref{eq:def-heat-kernel}, \eqref{eq:zeta-heat-kernel} implies that the zeta function is also uniquely defined and does not depend on the basis of eigenvectors.
Then, one has
\begin{equation}
	\tilde \zeta(s, x, y)
		= \sum_{n \neq 0} \frac{\slashed \nabla \Psi_n(x) \big( \slashed \nabla \Psi_n(y) \big)^\dagger}{\Lambda_n^{(0)} \Lambda_n^{s}}
		= \sum_{n \neq 0} \frac{\slashed \nabla \Psi_n(x) \big( \slashed \nabla \Psi_n(y) \big)^\dagger}{\Lambda_n^{s}(\Lambda_n - m^2)}.
\end{equation}
In order to make contact with the term above, we would like to get rid of the $m^2$ in the denominator.
To achieve this, we remark that
\begin{equation}
	\frac{1}{(\Lambda_n - m^2)} - \frac{m^2}{\Lambda_n (\Lambda_n - m^2)}
		= \frac{1}{\Lambda_n}.
\end{equation}
We find the second term of the LHS by shifting $s$ by $1$ and multiplying by $m^2$ in the expression of $\tilde \zeta(s, x, y)$ above.
As a consequence, we have:
\begin{equation}
	\tilde \zeta(s, x, y) - m^2 \tilde \zeta(s+1, x, y)
		= \sum_{n \neq 0} \frac{\slashed \nabla \Psi_n(x) \big( \slashed \nabla \Psi_n(y) \big)^\dagger}{\Lambda_n^{s+1}}.
\end{equation}

Next, we need to evaluate the other terms in \eqref{eq:third-term-variation-zeta}.
We expect the second term to be related to $\Delta \tilde \zeta$, since we obtain it by distributing one derivative on each spinor inside $\tilde \zeta$:
\begin{equation}
	\label{eq:laplacian-zeta}
	\begin{aligned}
	\tr_D \Delta \tilde \zeta(s, x, x)
		&
		= \tr_D \nabla^\mu \nabla_\mu \tilde \zeta(s + 1, x, x)
		= \tr_D \nabla^\mu \nabla_\mu \sum_{n \neq 0} \frac{\Psi_n(x) \Psi_n(x)^\dagger}{\Lambda_n^s}
		\\ & \hspace{-2cm}
		= \tr_D \sum_{n \neq 0} \frac{1}{\Lambda_n^s} \Big(
			\big( \Delta \Psi_n(x) \big) \Psi_n(x)^\dagger
			+ \Psi_n(x) \Psi_n(x)^\dagger \overleftarrow{\Delta}
			+ 2 \nabla_\mu \Psi_n(x) \big) \Psi_n(x)^\dagger \overleftarrow{\nabla}^\mu
			\Big)
		\\ & \hspace{-2cm}
		= \tr_D \sum_{n \neq 0} \frac{1}{\Lambda_n^s} \Big(
			\big( \Delta \Psi_n(x) \big) \Psi_n(x)^\dagger
			+ \Psi_n(x) \big( \Delta \Psi_n(x) \big)^\dagger
			+ 2 \nabla_\mu \Psi_n(x) \big) \big(\nabla^\mu \Psi_n(x)\big)^\dagger
			\Big).
	\end{aligned}
\end{equation}
In the last line, all covariant derivatives and Laplacians act on spinors.
Note that there is no ambiguity on the location of the matrices thanks to the presence of the trace.
We can then replace $\Delta$ using \eqref{eq:operator-D2}, before combining with the third term from \eqref{eq:third-term-variation-zeta} in $R$ and acting with $D^2$ on the eigenstates:
\begin{equation}
	\begin{aligned}
	2 \tr_D \sum_{n \neq 0} \frac{\nabla_\mu \Psi_n(x) \big) \big(\nabla^\mu \Psi_n(x)\big)^\dagger}{\Lambda_n^s}
		&
		= \tr_D \Delta \tilde \zeta(s, x, x)
			+ 2 \tr_D \tilde \zeta(s - 1, x, x)
			\\ & \qquad
			- 2 \left( \frac{R}{4} + m^2 \right) \tr_D \tilde \zeta(s, x, x).
	\end{aligned}
\end{equation}
Putting all the pieces together with $s \to s + 1$ gives the expression for $I_3$ in \eqref{eq:I-values}.

Note that $I_3 + I_4/2 = 0$, such that the total variation of the zeta function is:
\begin{equation}
	\delta\tilde\zeta(s) = 2s \int \dd^2 x \sqrt g \, \delta\sigma(x) \tr_D \tilde\zeta(s, x, x)
	-2 m^2 s \int \dd^2 x \sqrt g \, \delta\sigma(x) \tr_D \tilde \zeta(s+1, x, x)
\end{equation}
while its derivative is
\begin{align}
	\delta\tilde\zeta'(s) &= 2s \int \dd^2 x \sqrt g \, \delta\sigma(x) \tr_D \tilde \zeta'(s, x, x) + 2 \int \dd^2 x \sqrt g \, \delta\sigma(x) \tr_D \tilde \zeta(s, x, x) \nonumber \\
	&-2m^2 \left(s \int \dd^2 x \sqrt g \, \delta\sigma(x) \tr_D \tilde \zeta'(s+1, x, x) + \int \dd^2 x \sqrt g \, \delta\sigma(x) \tr_D \tilde \zeta(s+1, x, x)\right).
\end{align}

To compute the variation of the gravitational action, one needs the value of $\delta\tilde\zeta$ and $\delta\tilde\zeta'$ at $s = 0$.
They are given by
\begin{equation}
	\delta\tilde\zeta(0) = -2m^2 \lim_{s \to 0} s \int \dd^2 x \sqrt g \, \delta\sigma(x) \tr_D \tilde \zeta(s+1, x, x) = - \frac{m^2}{\pi} \int \dd^2 x \sqrt g \, \delta\sigma(x)
\end{equation}
and
\begin{align}
	\label{eq:variation-derivative-delta-zeta-0}
	\delta\tilde\zeta'(0) &= 2 \int \dd^2 x \sqrt g \, \delta\sigma(x) \tr_D \tilde \zeta(0, x, x) \nonumber \\
	&\quad - 2m^2 \lim_{s \to 0}\left[ \left(1 + s\frac{\dd}{\dd s} \right) \int \dd^2 x \sqrt g \, \delta\sigma(x) \tr_D \tilde \zeta(s+1, x, x)\right].
\end{align}
We recover the same formula as in \cite{Bilal:2017:2DQuantumGravity} for the case of the massive scalar field.

In the massive case, to compute the first term in \eqref{eq:variation-derivative-delta-zeta-0}, we use the fact that
\begin{equation}
	\tilde \zeta(0, x, x)
		= \zeta(0, x, x) - P(x, x)
		= \frac{a_1(x, x)}{4 \pi} - P(x, x)
\end{equation}
such that
\begin{equation}
	\begin{aligned}
		\int \dd^2 x \sqrt g \, \delta\sigma \, \tr_D \tilde \zeta(0, x, x)
			= &- \frac{1}{24 \pi} \int \dd^2 x \sqrt g \, \delta\sigma(x) R(x)
			- \int \dd^2 x \sqrt g \, \delta\sigma(x) \tr_D P(x, x)
			\\
			&- \frac{m^2}{2\pi} \int \dd^2 x \sqrt g \, \delta\sigma(x).
	\end{aligned}
\end{equation}
To compute the second term, we recognize $\zeta_{\text{reg}}$ from \eqref{eq:G-zeta}:
\begin{equation}
	\lim_{s \to 0}\left[ \left(1 + s\frac{\dd}{\dd s} \right) \int \dd^2 x \sqrt g \, \delta\sigma(x) \tr_D \tilde \zeta(s+1, x, x)\right]
		= \int \dd^2 x \sqrt g \, \delta\sigma \, \tilde \zeta_{\text{reg}}(1, x, x).
\end{equation}
Putting the pieces together, we find that
\begin{equation}
	\begin{aligned}
	\delta \zeta'(0)
		=
			&
			- \frac{1}{12 \pi} \int \dd^2 x \sqrt g \, \delta\sigma(x) R(x)
			- 2 \int \dd^2 x \sqrt g \, \delta\sigma(x) \tr_D P(x, x)
			\\ &
			- \frac{m^2}{\pi} \int \dd^2 x \sqrt g \, \delta\sigma(x)
			- 2 m^2 \int \dd^2 x \sqrt g \, \delta\sigma(x) \tr_D \tilde \zeta_{\text{reg}}(1, x, x).
	\end{aligned}
\end{equation}

The first term is recognized to be the variation of the Liouville action
\begin{equation}
	\delta S_L
		= \frac{1}{4 \pi} \int \dd^2 x \sqrt g \, \delta\sigma(x) R(x).
\end{equation}
The second term corresponds to the variation of the zero-mode normalization matrix \eqref{eq:variation-kappa}.
If there is a single constant zero-mode, like for the flat torus \eqref{eq:torus-kappa-proj}, this term is related to the variation of the area:
\begin{equation}
	\int \dd^2 x \sqrt g \, \delta\sigma(x) \tr_D P(x, x)
		= \delta \ln \det \kappa
		= - 2 \frac{\delta A}{A}.
\end{equation}
The third term contains is related to the variation of the Mabuchi action~\cite{Ferrari:2012:GravitationalActionsTwo}:
\begin{equation}
	\delta S_M
		\propto \int \dd^2 x \sqrt{g} \, \delta \sigma + \cdots
\end{equation}

The infinitesimal variation of the gravitational action is
\begin{equation}
	\label{expression-delta-Sgrav}
	\begin{aligned}
	\delta S_{\text{grav}}
		= &-\frac{1}{12} \, \delta S_L
			- \frac{1}{2} \, \delta \ln \det \kappa \\
			&- \frac{m^2}{2} \int \dd^2 x \sqrt g \, \delta\sigma(x) \tr_D \left(\tilde \zeta_{\text{reg}}(1, x, x) + \frac{\ln \mu^2}{4\pi} + \frac{1}{4\pi}\right).
	\end{aligned}
\end{equation}
We can express this in terms of the regularized Green's function by noting that
\begin{equation}
	\tilde \zeta_{\text{reg}}(1, x, x) + \frac{\ln\mu^2}{4\pi}
		= \tilde G_\zeta(x)
\end{equation}
so that
\begin{subequations}
\begin{align}
	\delta S_{\text{grav}}
		&= - \frac{1}{12} \, \delta S_L
			- \frac{1}{2} \, \delta \ln \det \kappa
			- \frac{m^2}{2} \int \dd^2 x \sqrt g \, \delta\sigma(x) \tr_D \left( \tilde G_\zeta(x) + \frac{1}{4\pi} \right) \\
		&= - \frac{1}{12} \, \delta S_L - \frac{m^2}{2} \int \dd^2 x \sqrt g \, \delta\sigma(x) \tr_D \left( G_\zeta(x) + \frac{1}{4\pi} \right).
\end{align}
\end{subequations}

Now we want to express $\delta S_{\text{grav}}$ as the variation of some functional. For this, we note that
\begin{equation}
	\int \dd^2 x \sqrt g \, \delta \sigma \tr_D G_\zeta(x) = \frac{1}{2} \, \delta\left(\int \dd^2 x \sqrt g \, \tr_D G_\zeta (x)\right) - \frac{1}{2} \int \dd^2 x \sqrt g \tr_D \delta G_\zeta(x).
\end{equation}
Using \eqref{eq:deltaG} and \eqref{eq:var-G-zeta}, computations similar to the previous steps lead to
\begin{equation}
	\int \dd^2 x \sqrt g \tr_D \delta G_\zeta(x) = - 2m^2 \int \dd^2 x \sqrt g \, \delta \sigma(x) \tr_D \zeta(2, x, x) + \frac{1}{2\pi} \int \dd^2 x \sqrt g \, \delta \sigma(x).
\end{equation}
This gives:
\begin{equation}
	\label{expression-deltaSgrav-2}
	\delta S_{\text{grav}} = - \frac{1}{12} \, \delta S_L - \frac{m^2}{4} \, \delta\left(\int \dd^2 x \sqrt g \, \tr_D G_\zeta(x)\right) - \frac{m^4}{2} \int \dd^2 x \sqrt g \, \delta \sigma(x) \tr_D \zeta(2, x, x).
\end{equation}
To go further, we show that
\begin{equation}
	\label{eq:Kzeta2}
	\begin{aligned}
		m^4 \int \dd^2 x \sqrt g \, \delta \sigma(x) \tr_D \zeta(2, x, x) = &\ \frac{1}{4} \int_0^\infty \dd t \, \frac{\e^{m^2 t} - m^2 t -1}{t} \, \delta \tilde K(t) \\
		&+ \int \dd^2 x \sqrt g \, \delta \sigma(x) \tr_D P(x, x)
	\end{aligned}
\end{equation}
in \Cref{app:proof-identities-Kzeta}.
Reporting in \eqref{expression-deltaSgrav-2}, we get
\begin{equation}
	\label{expression-deltaSgrav-final}
	\begin{aligned}
	\delta S_{\text{grav}}
		= &- \frac{1}{12} \, \delta S_L
		- \frac{1}{2} \, \delta \ln \det \kappa
		- \frac{m^2}{4} \, \delta\left(\int \dd^2 x \sqrt g \, \tr_D G_\zeta(x)\right)
		\\
		&- \frac{1}{8} \, \delta \left(\int_0^\infty \frac{\dd t}{t} \, \big(\e^{m^2 t} - m^2 t - 1 \big) \, \tilde K(t)\right).
	\end{aligned}
\end{equation}

Because \eqref{expression-deltaSgrav-final} is a total variation, it can be integrated:
\begin{equation}
	\begin{aligned}
	S_{\text{grav}}[\hat g, g]
		= &- \frac{1}{12} \, S_L[\hat g, g]
		- \frac{1}{2} \, \ln \det \frac{\kappa[g]}{\kappa[\hat g]}
		\\
		&- \frac{m^2}{4} \int \dd^2 x \, \tr_D \big( \sqrt g \, G_{g,\zeta}(x) - \sqrt{\hat g} \, G_{\hat g, \zeta}(x) \big)
		\\
		&- \frac{1}{8} \, \int_0^\infty \frac{\dd t}{t} \, \big(\e^{m^2 t} - m^2 t - 1 \big) \, \big( \tilde K_g(t) - \tilde K_{\hat g}(t) \big).
	\end{aligned}
\end{equation}
A last simplification can be achieved by noting that the difference between $G$ and $\tilde G$ is the same constant, proportional to the number of zero-modes, using \eqref{eq:G-Gtilde} and \eqref{eq:trace-projector}.
This, it cancels and one can replace $G$ by $\tilde G$
\begin{equation}
	\label{expression-Sgrav-final}
	\begin{aligned}
	S_{\text{grav}}[\hat g, g]
		= &- \frac{1}{12} \, S_L[\hat g, g]
		- \frac{1}{2} \, \ln \det \frac{\kappa[g]}{\kappa[\hat g]}
		\\
		&- \frac{m^2}{4} \int \dd^2 x \, \tr_D \big( \sqrt g \, \tilde G_{g,\zeta}(x) - \sqrt{\hat g} \, \tilde G_{\hat g, \zeta}(x) \big)
		\\
		&- \frac{1}{8} \, \int_0^\infty \frac{\dd t}{t} \, \big(\e^{m^2 t} - m^2 t - 1 \big) \, \big( \tilde K_g(t) - \tilde K_{\hat g}(t) \big).
	\end{aligned}
\end{equation}
The first two terms, which would be the only ones in the massless case, arise from the conformal anomaly and agree with the standard result, giving the Liouville action for a CFT with central charge $c = 1/2$~\cite{Blau:1989:DeterminantsDiracOperators}.

Note that $\ln \det \kappa[g] / \kappa[\hat g]$ can be rewritten as a determinant inside the path integral: it takes into account the redefinition of the modes that would become zero-modes in the massless limit.
To make contact with the scalar field computation of~\cite{Bilal:2017:2DQuantumGravity}, this expression gives the correct factor $\ln(A / \hat A)$ using any choice of normalization (see \cref{ft:scalar-zero-mode}).
When there is a single constant zero-mode, such as for the flat torus \eqref{eq:torus-kappa-proj}, we get:
\begin{equation}
	- \frac{1}{2} \, \ln \det \frac{\kappa[g]}{\kappa[\hat g]}
		= \ln \frac{A}{\hat A}.
\end{equation}

The expression \eqref{expression-Sgrav-final} coincides with the result for the scalar field from~\cite{Bilal:2017:2DQuantumGravity} up to factors $\pm 1/2$.
One can guess that the different signs come from the statistics of the fermion, and the factor $1/2$ from the central charge.
Studying new models would be useful for checking this last guess.
Moreover, it would be very interesting to interpret these facts in view of supersymmetry.

\subsection{Small mass expansion}

The goal of this section is to study the small mass expansion of \eqref{expression-Sgrav-final}.

First, one notes that
\begin{equation}
	\int_0^\infty \frac{\dd t}{t} \, \big(\e^{m^2 t} - m^2 t - 1 \big) \, K(t) = O(m^4).
\end{equation}
Then, because of the factor $m^2$ in front of $\tilde G_\zeta$, it is sufficient to consider the limit $m \to 0$ of $\tilde G$, which is well-defined since we removed the zero-modes.
As a consequence, the action \eqref{expression-Sgrav-final} reduces to
\begin{equation}
	\label{Sgrav-small-mass}
	\begin{aligned}
	S_{\text{grav}}[\hat g, g]
		= &- \frac{1}{12} \, S_L[\hat g, g]
		- \frac{1}{2} \, \ln \det \frac{\kappa[g]}{\kappa[\hat g]}
		\\
		&- \frac{m^2}{4} \int \dd^2 x \, \tr_D \big( \sqrt g \, G^{(0)}_{g,\zeta}(x) - \sqrt{\hat g} \, G^{(0)}_{\hat g, \zeta}(x) \big)
		+ O(m^4).
	\end{aligned}
\end{equation}
The symbol $O(m^4)$ is omitted from now on.
It can be useful to work directly with the variation
\begin{subequations}
\label{eq:variation-Sgrav-small-mass}
\begin{align}
	\delta S_{\text{grav}}
		&= - \frac{1}{12} \, \delta S_L
			- \frac{1}{2} \, \delta \ln \det \kappa
			- \frac{m^2}{4} \, \delta\left(\int \dd^2 x \sqrt g \, \tr_D \tilde G_\zeta^{(0)}(x)\right)
		\\
		&= - \frac{1}{12} \, \delta S_L
			- \frac{1}{2} \, \delta \ln \det \kappa
			- \frac{m^2}{2} \int \dd^2 x \sqrt g \, \delta\sigma(x) \tr_D \left( \tilde G_\zeta^{(0)}(x) + \frac{1}{4\pi} \right)
		\\
		&= - \frac{1}{12} \, \delta S_L
			- \frac{m^2}{2} \int \dd^2 x \sqrt g \, \delta\sigma(x) \tr_D \left( G_\zeta^{(0)}(x) + \frac{1}{4\pi} \right).
\end{align}
\end{subequations}

If we directly follow the scalar case, then one can introduce the averaged integrated Green function
\begin{equation}
	\label{eq:integrated-G}
	\Psi_G[g]
		= \frac{1}{A} \int \dd^2 x \sqrt{g} \, \tr_D G_{g,\zeta}^{(0)}(x).
\end{equation}
In this case, the above expression becomes
\begin{equation}
	\label{eq:Sgrav-small-mass}
	S_{\text{grav}}[\hat g, g]
		= - \frac{1}{12} \, S_L[\hat g, g]
			- \frac{1}{2} \, \ln \det \frac{\kappa[g]}{\kappa[\hat g]}
			- \frac{m^2 A}{4} \, (\Psi_G[g] - \Psi_G[\hat g]) + \frac{m^2}{4} (A - \hat A) \Psi_G[\hat g].
\end{equation}
The last term is independent of $g$ and contributes to the cosmological constant.
The second term becomes the usual $\ln A / \hat A$ when there is a single constant zero-mode, see \eqref{eq:torus-kappa-proj}.

Note that the variation of the last term in \eqref{eq:variation-Sgrav-small-mass} can also be integrated directly to $\sigma \e^{2 \sigma}$, which shows that the $O(m^2)$ term should contain the Mabuchi action~\cite{Ferrari:2012:GravitationalActionsTwo,Bilal:2021:EffectiveGravitationalAction}:
\begin{equation}
	S_M[\hat g, \sigma]
		= \frac{4}{A} \int \dd^2 x \, \sqrt{\hat g} \, \sigma \e^{2 \sigma}
			+ \cdots
\end{equation}
Note that the coefficient is $- m^2 A / 16 \pi$, to be compared with a factor $+ m^2 A / 16 \pi$ for a scalar field~\cite{Ferrari:2012:GravitationalActionsTwo,Bilal:2021:EffectiveGravitationalAction}.
As noted above, this may be related to supersymmetry.

We refrain from providing explicit formulas for \eqref{eq:Sgrav-small-mass} in terms of the $\hat g$-dependent Green functions, since they are not particularly insightful.
However, we highlight the procedure to compute the finite variation of $\Psi_G[g]$.
From \eqref{eq:integrated-G}, we see that we need the finite variation of the regularized Green function $G_{g,\zeta}(x)$, which can be obtained from the variation of the Green function $G(x,y)$.
The latter can be found from the finite transformation of the Green function $\widetilde S^{(0)}(x, y)$ using~\cite[eq.~(3.32)]{Dettki:1993:GeneralizedGaugedThirring}:
\begin{equation}
	\label{eq:weyl-tilde-Sxy}
	\begin{aligned}
	\widetilde S^{(0)}_{g,xy}
		&
		=
			\e^{-\frac{\phi_x}{2}}
				\widetilde S^{(0)}_{\hat g,xy}
				\e^{-\frac{\phi_y}{2}}
			+ \int \dd^2 w \sqrt{g} \int \dd^2 z \sqrt{g} \,
				P_{g,xw}
				\e^{-\frac{\phi_w}{2}}
				\widetilde S^{(0)}_{\hat g,wz}
				\e^{-\frac{\phi_z}{2}}
				P_{g,zy}
			\\ & \qquad
			- \int \dd^2 z \sqrt{g} \,
				P_{g,xz}
				\e^{-\frac{\phi_z}{2}}
				\widetilde S^{(0)}_{\hat g,zy}
				\e^{-\frac{\phi_y}{2}}
			- \int \dd^2 z \sqrt{g} \,
				\e^{-\frac{\phi_x}{2}}
				\widetilde S^{(0)}_{\hat g,xz}
				\e^{-\frac{\phi_z}{2}}
				P_{g,zy}
	\end{aligned}
\end{equation}
(to simplify the notations, we write the coordinate dependence as a subscript).
This is because the Green function must be orthogonal to the zero-modes, so we need $S \sim (1 - P_{\hat g}) \Sigma (1 - P_{\hat g})$, where $\Sigma = \e^{- (\sigma(x) + \sigma(y)) / 2} S_{\hat g}$ since the Dirac operator transforms covariantly under Weyl transformations.
This generalizes the formula found in~\cite{Steiner:2005:GeometricalMassIts, Doyle:2017:BlowingBubblesTorus, Okikiolu:2007:ExtremalsLogarithmicHardyLittlewoodSobolev, Okikiolu:2009:NegativeMassTheorem, Morpurgo:1996:LogarithmicHardyLittlewoodSobolevInequality} for the massless scalar.
The projector \eqref{eq:proj-zero-mode} can be written in terms of the $\hat g$ zero-modes as:
\begin{equation}
	P_g(x, y)
		= \e^{- \frac{\sigma(x)}{2} - \frac{\sigma(y)}{2}}
			\sum_{i,j} \hat \Psi_{0, i}(x) \kappa^{ij}[g] \hat \Psi^\dagger_{0, j}(y),
\end{equation}
but there is no simple formula in terms of $P_{\hat g}$ because of the presence of $\kappa[g]$.

\section*{Acknowledgements}

We would like to thank Andreas Wipf for useful discussions.

This project has received funding from the European Union's Horizon 2020 research and innovation program under the Marie Skłodowska-Curie grant agreement No 891169.
This work is supported by the National Science Foundation under Cooperative Agreement PHY-2019786 (The NSF AI Institute for Artificial Intelligence and Fundamental Interactions, \url{http://iaifi.org/}).

\appendix

\section{Formulas and conventions}
\label{sec:conventions}

Greek indices refer to the curved frame and Latin indices ($a, b$, etc.) to the local flat frame.
Explicit indices are denoted by letters in the first case, $\mu = (t, x)$, and numbers in the second case, $a = 0, 1$.
We globally follow the conventions of~\cite{Freedman:2012:Supergravity, Blumenhagen:2014:BasicConceptsString}.

The metric of the two-dimensional compact curved manifold with Euclidean metric $g_{\mu\nu}$, where $\mu = 0, 1$.
Zweibeine $e^\mu_a$ are introduced to convert curved indices to local indices:
\begin{equation}
	\delta_{ab}
		= e^\mu_a e^\nu_b g_{\mu\nu}
		= \diag(1, 1).
\end{equation}
The antisymmetric Levi--Civita symbol $\epsilon_{ab}$ is normalized as
\begin{equation}
	\epsilon_{01}
		= \epsilon^{01} = 1.
\end{equation}

Complex coordinates are defined by
\begin{equation}
	z
		= x^0 + \I x^1, \qquad
	\bar z
		= x^0 - \I x^1
\end{equation}
such that
\begin{equation}
	\dd s^2
		= \dd z \dd \bar z,
	\qquad
	g_{zz}
		= g_{\bar z \bar z}
		= 0,
	\qquad
	g_{z \bar z}
		= \frac{1}{2}.
\end{equation}
The derivatives with respect to the complex coordinates are:
\begin{equation}
	\pd
		:= \pd_z
		= \frac{1}{2} (\pd_0 - \I \pd_1), \qquad
	\bar\pd
		:= \pd_{\bar z}
		= \frac{1}{2} (\pd_0 + \I \pd_1).
\end{equation}
The integration measure is
\begin{equation}
	\dd^2 z
		= 2 \dd^2 x,
	\qquad
	\dd^2 z
		:= \dd z \dd \bar z,
	\qquad
	\dd^2 x
		:= \dd x^0 \dd x^1.
\end{equation}

\subsection{Covariant derivatives}

The covariant derivative is denoted by
\begin{equation}
	\nabla_\mu
		:= \pd_\mu + \Gamma_\mu,
\end{equation}
where $\Gamma_\mu$ is the connection which contains a combination of the Levi--Civita $\tensor{\Gamma}{_{\mu\nu}^{\rho}}$ and spin $\tensor{\omega}{_{\mu a}^{b}}$ connections depending on the object it acts on.
The Laplacian is defined by
\begin{equation}
	\Delta
		:= g^{\mu\nu} \nabla_\mu \nabla_\nu.
\end{equation}

The covariant derivative of a fermion reads
\begin{equation}
	\label{eq:grad-fermion}
	\nabla_\mu \Psi
		= \left(\pd_\mu + \frac{1}{4}\, \omega_{\mu ab} \gamma^{ab} \right) \Psi
		= \left(\pd_\mu - \frac{\I}{4}\, \omega_{\mu} \gamma_* \right) \Psi,
\end{equation}
where $\gamma_*$ is defined in \eqref{eq:gamma-star} and:
\begin{equation}
	\omega_{\mu}
		:= \omega_{\mu ab} \epsilon^{ab}.
\end{equation}
The covariant derivative of gamma matrices vanishes:
\begin{equation}
	\label{eq:grad-gamma}
	\nabla_\mu \gamma^\nu
		= 0.
\end{equation}
The square of the Dirac operator $\slashed\nabla$ is:
\begin{equation}
	\label{eq:dirac-squared}
	\slashed\nabla^2
		= \Delta - \frac{R}{4},
\end{equation}
where the spinorial Laplacian is
\begin{equation}
	\label{def:spinor-Laplacian}
	\Delta \Psi
		= g^{\mu\nu} \left[
			\left(\partial_\mu + \frac{1}{4} \omega_{\mu a b} \gamma^{ab} \right) \nabla_\nu \Psi
			- \Gamma_{\mu\nu}^\rho \nabla_\rho \Psi
			\right].
\end{equation}
We have used:
\begin{equation}
	\label{eq:antisym-deriv}
	\gamma^{\mu\nu} \com{\nabla_\mu}{\nabla_\nu}
		= - \frac{R}{2}.
\end{equation}

The Dirac operator together with a mass term is denoted by
\begin{equation}
	D
		:= \I \slashed\nabla + m \gamma_*.
\end{equation}
The square of this operator reads
\begin{equation}
	D^2
		= - \Delta + \frac{R}{4} + m^2,
\end{equation}
which follows from $\gamma_*^2 = 1$ and $\com{\gamma^\mu}{\gamma_*} = 0$.
Due to the additional $\gamma_*$ in the mass term, and contrary to the usual case, it is not necessary to conjugate the $D$ operator by $\gamma_*$ in order to obtain the RHS.

\subsection{Conformal variations}

The conformal variation of the metric reads
\begin{equation}
	g_{\mu\nu}
		= \e^{2\sigma} \hat g_{\mu\nu},
\end{equation}
which implies:
\begin{equation}
	e_\mu^a
		= \e^\sigma \, \hat e_\mu^a.
\end{equation}
Then, the affine et spin connections transform as:
\begin{align}
	\Gamma^\mu_{\nu\rho}
		= \hat \Gamma^\mu_{\nu\rho}
			+ \delta^\mu_\nu \partial_\rho \sigma
			+  \delta^\mu_\rho \partial_\nu \sigma
			- \hat g_{\nu\rho} \, \partial^\mu \sigma,
	\\
	\omega_{\mu ab}
		= \hat \omega_{\mu ab}
		+ \hat e_{\mu a} \, \hat e^\nu_b \, \partial_\nu \sigma
		- \hat e_{\mu b} \, \hat e^\nu_a \, \partial_\nu \sigma.
\end{align}
This gives the expression for the covariant derivative for a spinor:
\begin{equation}
	\nabla_\mu \Psi
		= \hat \nabla_\mu \Psi
			+ \frac{1}{2}\, \hat e_{\mu a} \, \hat e^\nu_b \, \partial_\nu \sigma \, \gamma^{ab} \, \Psi,
\end{equation}
such that
\begin{equation}
	\label{eq:transformation-slashed-nabla}
	\slashed \nabla
		= \e^{-\sigma} \left(
			\hat{\slashed\nabla}
			+ \frac{1}{2} \, \partial_\mu \sigma \hat \gamma^\mu
			\right).
\end{equation}

The transformation of the spinor Laplacian $\Delta$ is:
\begin{equation}
	\Delta
		= \e^{-2\sigma} \left(
			\hat \Delta +
			\partial_\nu \sigma \, \hat \gamma^{\mu\nu} \, \hat \nabla_\mu
			- \frac{1}{4} \, \partial_\mu \sigma \, \partial^\mu \sigma
			\right).
\end{equation}
Finally, from the transformation of the Ricci scalar
\begin{equation}
	\label{eq:variationR}
	R
		= \e^{-2\sigma} \big(
			\hat R
			- 2 \, \hat \Delta \sigma
			\big),
\end{equation}
we find:
\begin{equation}
	\slashed \nabla^2
		= \e^{-2\sigma} \left(
			- \hat \Delta
			+ \frac{1}{4} \, \hat R
			- \partial_\nu \sigma \, \hat \gamma^{\mu\nu} \, \hat \nabla_\mu
			+ \frac{1}{4} \, \partial_\mu \sigma \, \partial^\mu \sigma
			- \frac{1}{2}  \, \hat \Delta \sigma
			\right).
\end{equation}

From the last formula, we can deduce the infinitesimal variation of the operator $D^2$:
\begin{equation}
	\label{eq:variation-D2}
	\delta D^2
		= - 2 \, \delta\sigma \left( - \Delta + \frac{R}{4} \right)
			- \partial_\nu (\delta \sigma) \, \gamma^{\mu\nu} \nabla_\mu
			- \frac{1}{2} \, \Delta (\delta \sigma).
\end{equation}

\section{Two-dimensional spinors}
\label{sec:2d-fermions}

This appendix summarizes the main properties of spinors and gamma matrices.
For general references, the reader is referred to~\cites[app.~7.5, 8.5]{Blumenhagen:2014:BasicConceptsString}[chap.~2, 3]{Freedman:2012:Supergravity}{VanProeyen:1999:ToolsSupersymmetry}[sec.~13.2]{Wipf:2016:IntroductionSupersymmetry}.
In this appendix, we will work with local indices.

\subsection{Clifford algebra and gamma matrices}

The two-dimensional identity is denoted by
\begin{equation}
	\mathrm I_2 =
		\begin{pmatrix}
			1 & 0 \\
			0 & 1
		\end{pmatrix},
\end{equation}
or with $1$ when no confusion is possible.

The $\group{SO}(2)$ Clifford algebra is generated by the two gamma matrices $\gamma^a$ satisfying the anticommutation relation
\begin{equation}
	\anticom{\gamma^{a}}{\gamma^{b}}
		= 2 \delta^{ab}.
\end{equation}
Both matrices are taken to be unitary, and as a consequence Hermitian
\begin{equation}
	\adj{(\gamma^{a})}
		= \gamma^{a}.
\end{equation}

The last element of the algebra corresponds to the antisymmetric product
\begin{equation}
	\gamma^{ab}
		= \frac{1}{2} \, \com{\gamma^a}{\gamma^b}
		= - \I \, \epsilon^{ab} \gamma_*,
\end{equation}
which is proportional to the chirality matrix
\begin{equation}
	\label{eq:gamma-star}
	\gamma_*
		= \I \, \gamma^0 \gamma^1
		= \frac{\I}{2} \, \epsilon_{ab} \gamma^{ab}.
\end{equation}
It corresponds to the generator of the $\group{SO}(2)$ group
\begin{equation}
	M^{ab}
		= \frac{\I}{2} \, \gamma^{ab}
		= \epsilon^{ab} \gamma_*.
\end{equation}
The chirality matrix is Hermitian, unitary and anticommutes will other gamma matrices
\begin{equation}
	\gamma_*^2
		= 1,
	\qquad
	\adj{(\gamma_*)}
		= \gamma_*,
	\qquad
	\anticom{\gamma_*}{\gamma^a}
		= 0.
\end{equation}

Some useful identities are
\begin{subequations}
\begin{gather}
	\label{eq:gamma-sandwich-gamma}
	\gamma^a \gamma_b \gamma_a
		= 0 \\
	\gamma^a \gamma^b
		= \eta^{ab} - \I \, \varepsilon^{ab} \gamma_* \\
	\gamma^{ab} \gamma_{bc}
		= \delta^a_c \\
	\gamma_* \gamma^a
		= \varepsilon^{ab} \gamma_b \\
	\com{\gamma^{ab}}{\gamma^c}
		= 2 \varepsilon^{ab} \varepsilon^{cd} \gamma_d.
\end{gather}
\end{subequations}

\subsection{Dirac and Majorana spinors}

A Dirac spinor $\Psi$ is a $2$-dimensional complex vector with anticommuting components that forms a reducible representation of the Clifford algebra.
Such a spinor transforms under a Lorentz transformation as
\begin{equation}
	\delta \Psi
		= - \frac{1}{4} \, \lambda_{ab} M^{ab} \Psi
		= - \frac{\lambda}{4} \,  \gamma_* \Psi,
	\qquad
	\lambda := \lambda_{ab} \epsilon^{ab}.
\end{equation}

The Dirac conjugation corresponds to Hermitian conjugation
\begin{equation}
	\bar{\Psi}
		= \adj{\Psi}.
\end{equation}
This object transforms as
\begin{equation}
	\delta \bar{\Psi}
		= \frac{\lambda}{4} \, \bar{\Psi} \gamma_*
\end{equation}
such that $\bar\Psi \Psi$ is a scalar.

Introducing the charge conjugation matrix $C$ such that
\begin{equation}
	\label{eq:charge-conj-def}
	\conj{(\gamma^\mu)}
		= C \gamma^\mu C^{-1},
	\qquad
	(\gamma^\mu)^t
		= C \gamma^\mu C^{-1},
\end{equation}
the charge conjugated spinor and its Dirac conjugate (giving the Majorana conjugate) are defined by
\begin{equation}
	\Psi^c
		= C^{-1} \conj{\Psi},
	\qquad
	\bar{\Psi}^c
		= \Psi^t C.
\end{equation}
Both spinors transform respectively as $\Psi$ and $\bar\Psi$.
Note that $C$ is unitary and symmetric.

From a Dirac spinor, one can obtain two different irreducible representations: a Weyl (or chiral) spinor or a Majorana (or real) spinor.
The latter is given by the reality condition
\begin{equation}
	\Psi^c
		= \Psi
	\quad \Longrightarrow \quad
	\conj{\Psi}
		= C \Psi.
\end{equation}
This implies in particular that the Dirac and Majorana conjugations coincide.
A Weyl spinor is obtained from a Dirac spinor by projecting it on its positive or negative chirality
\begin{equation}
	\Psi
		= P_\pm \Psi,
	\qquad
	P_\pm
		= \frac{1}{2} (1 \pm \gamma_*).
\end{equation}
One should note that it is not possible to have Majorana--Weyl fermion in Euclidean signature (contrary to what happens in Lorentz signature).

Given two Majorana spinors $\Psi_1$ and $\Psi_2$, the Majorana flip relations read~\cite{Wipf:2016:IntroductionSupersymmetry}
\begin{equation}
	\label{eq:flip-majorana}
	\bar{\Psi}_1 \Psi_2
		= - \bar{\Psi}_2 \Psi_1,
	\qquad
	\bar{\Psi}_1 \gamma_\mu \Psi_2
		= - \bar{\Psi}_2 \gamma_\mu \Psi_1,
	\qquad
	\bar{\Psi}_1 \gamma_* \Psi_2
		= \bar{\Psi}_2 \gamma_* \Psi_1.
\end{equation}

\subsection{Gamma matrix representations}

\subsubsection{Majorana basis}

In the Majorana representation, the Dirac matrices read
\begin{equation}
	\gamma^0
		= \sigma_1
		=
		\begin{pmatrix}
			0 & 1 \\
			1 & 0
		\end{pmatrix},
	\qquad
	\gamma^1
		= \sigma_3
		=
		\begin{pmatrix}
			1 & 0 \\
			0 & -1
		\end{pmatrix},
	\qquad
	\gamma_*
		= \sigma_2
		=
		\begin{pmatrix}
			0 & - \I \\
			\I & 0
		\end{pmatrix},
\end{equation}
which implies that the charge conjugation is the identity
\begin{equation}
	C
		= 1.
\end{equation}

In this basis, a Majorana spinor has real components
\begin{equation}
	\Psi =
	\begin{pmatrix}
		\psi_1 \\ \psi_2
	\end{pmatrix},
\qquad
	\conj{\psi_1}
		= \psi_1, \quad
	\conj{\psi_2}
		= \psi_2,
\end{equation}
The scalar bilinears are
\begin{equation}
	\bar\Psi \Psi
		= (\psi_1)^2 + (\psi_2)^2
		= 0,
	\qquad
	\bar\Psi \gamma_* \Psi
		= - 2 \I\, \psi_1 \psi_2,
\end{equation}
while the kinetic operator reads
\begin{equation}
	\I \, \gamma^\mu \pd_\mu
		= \I
		\begin{pmatrix}
			\pd_1 & \pd_0 \\
			\pd_0 & - \pd_1
		\end{pmatrix}.
\end{equation}

\subsubsection{Weyl basis}

In the Majorana representation, the Dirac matrices read
\begin{equation}
	\label{eq:weyl-basis}
	\Gamma^0
		= \sigma_2
		=
		\begin{pmatrix}
			0 & - \I \\
			\I & 0
		\end{pmatrix},
	\qquad
	\Gamma^1
		= \sigma_1
		=
		\begin{pmatrix}
			0 & 1 \\
			1 & 0
		\end{pmatrix},
	\qquad
	\Gamma_*
		= \sigma_3
		=
		\begin{pmatrix}
			1 & 0 \\
			0 & - 1
		\end{pmatrix},
\end{equation}
which implies that the charge conjugation is:
\begin{equation}
	C
		= \Gamma^1.
\end{equation}

In this basis, a Majorana spinor has complex components conjugated to each other
\begin{equation}
	\Psi =
	\begin{pmatrix}
		\bar\psi \\ \psi
	\end{pmatrix},
\qquad
	\conj{\psi}
		= \bar\psi, \quad
	\conj{\bar\psi}
		= \psi.
\end{equation}
The relation with the Weyl components is
\begin{equation}
	\psi
		= \frac{1}{\sqrt{2}} (\psi_1 + \I \psi_2),
	\qquad
	\bar\psi
		= \frac{1}{\sqrt{2}} (\psi_1 - \I \psi_2).
\end{equation}
The scalar bilinears are
\begin{equation}
	\bar\Psi \Psi
		= \psi \bar\psi + \bar\psi \psi
		= 0,
	\qquad
	\bar\Psi \gamma_* \Psi
		= 2 \psi \bar\psi,
\end{equation}
while the kinetic operator reads
\begin{equation}
	\I \, \gamma^\mu \pd_\mu
		=
		\begin{pmatrix}
			0 & \bar\pd \\
			- \pd & 0
		\end{pmatrix}.
\end{equation}

\section{DDK ansatz}
\label{sec:ddk-ansatz}

In their seminal papers~\cite{David:1988:ConformalFieldTheories, Distler:1989:ConformalFieldTheory}, David, Distler and Kawai (DDK) addressed two questions of importance: the rewriting of the functional integral in terms of a free-field measure for the Liouville mode\footnotemark{} and the gravitational dressing of matter operator.
\footnotetext{%
    It would be particularly interesting to extend this analysis to the case of non-conformal matter, in particular for the case of the Liouville--Mabuchi gravity.
    We are grateful to E.\ D'Hoker for discussions on this topic.
}%
The latter has been used to provide an ansatz for the action associated to a CFT deformed by primary operators coupled to gravity or on a curved background (see~\cite{Seiberg:1990:NotesQuantumLiouville, Moore:1992:GravitationalPhaseTransitions, Eguchi:1993:C1LiouvilleTheory, Hsu:1993:GravitationalSineGordonModel, Schmidhuber:1993:ExactlyMarginalOperators, Ambjorn:1994:2DQuantumGravity, Schmidhuber:1994:RGFlow2d, Reuter:1996:WeylInvariantQuantizationLiouville, Reuter:1997:QuantumLiouvilleField, daCunha:2003:ClosedStringTachyon, Doyon:2004:IsingFieldTheory, Zamolodchikov:2002:ScalingLeeYangModel, Zamolodchikov:2005:PerturbedConformalField, Zamolodchikov:2006:MassiveMajoranaFermion, Martinec:2014:ModelingQuantumGravity} for a selection of references).
In practice, the ansatz is found by performing a conformal deformation of the gravitational action of the CFT with the additional constraint that these deformations have a conformal weight equal to one.

The first point received strong supports by explicit computations~\cite{Mavromatos:1989:RegularizingFunctionalIntegral, DHoker:1990:2DQuantumGravity, DHoker:1991:EquivalenceLiouvilleTheory} and indirect approaches showing the consistency of the theory with a free-field measure~\cite{Teschner:2001:LiouvilleTheoryRevisited, Ribault:2022:ConformalFieldTheory, Ribault:2015:LiouvilleTheoryCentral, Rhodes:2016:LectureNotesGaussian, Kupiainen:2016:ConstructiveLiouvilleConformal}.
However, the validity of the second is more questionable.
We will discuss this and set it in the context of this paper in \Cref{sec:ddk:discussion} after giving a general review of the DDK ansatz in \Cref{sec:ddk:review}.

\subsection{Review of the DDK ansatz}
\label{sec:ddk:review}

The matter theory is described by the action $S_{\text{cft}}$ of a CFT deformed by primary operators
\begin{subequations}
\begin{gather}
    S_m[g, \psi] = S_{\text{cft}}[g, \psi] + S_p[g, \psi], \\
    S_p[g, \psi] = \sum_i \lambda^i \int \dd^2 x \sqrt{g} \, \mc O_i(\psi),
\end{gather}
\end{subequations}
where $\psi$ denotes collectively the fields, $\lambda^i$ the coupling constants, and $\mc O_i(\psi)$ is a set of primary operators\footnotemark{} with conformal weight $h_i$ and built from the fields $\psi$.
To simplify the discussion, we consider only spinless fields such that the conformal dimension is $2 h_i$.
\footnotetext{%
    While the operator $\mc O_i$ can depend explicitly on the metric there is also an implicit metric dependence which is due to the regularization needed to remove self-contraction.
    This is discussed for example in~\cites{DHoker:1991:EquivalenceLiouvilleTheory, Dorn:1992:AnalysisAllDimensionful, Kawai:1993:UltravioletStableFixed}[sec.~3.6]{Polchinski:2005:StringTheory-1}{Bautista:2016:QuantumCosmologyTwo} for the case of the scalar field.
}%
Hence, a Weyl transformation acts as
\begin{equation}
    \label{eq:weyl-g-O}
    g = \e^{2\omega} \hat g, \qquad
    \mc O_i = \e^{- 2 h_i \omega}\, \hat{\mc O}_i.
\end{equation}
The action $S_{\text{cft}}$ is conformally invariant on flat space $g = \delta$, and we assume it invariant under Weyl transformations (see~\cite{Iorio:1997:WeylGaugingConformalInvariance} for a discussion of this topic).
On the other hand, the action $S_p$ is not invariant if $h_i \neq 1$ for at least one operator.
The trace of the energy-momentum tensor for the perturbation is
\begin{equation}
    \label{eq:trace-pert}
    T^{(p)} = 4\pi \sum_i \lambda^i \mc O_i
\end{equation}
(in the absence of explicit metric dependence in the operators $\mc O_i$) and displays the breaking of the conformal invariance by the deformations.

The DDK ansatz proposes the total action in the conformal gauge to be given by
\begin{subequations}
\label{eq:ddk-ansatz}
\begin{gather}
    S_{\text{DDK}}[g, \sigma, \psi] = Q^2 \, S_L[\hat g, \sigma] + S_{\text{cft}}[\hat g, \psi] + S^{(p)}_{\text{DDK}}[\hat g, \sigma, \psi], \\
    \label{eq:ddk-action}
    S^{(p)}_{\text{DDK}}[\hat g, \sigma, \psi] = \sum_i \lambda^i \int \dd^2 x \sqrt{\hat g} \, \e^{2 a_i Q \sigma} \hat{\mc O}_i(\psi).
\end{gather}
\end{subequations}
We recall that $Q$ is related to the central charge by $6 Q^2 = 26 - c_m$ ($26$ arises from the ghost contribution).
The $a_i$ are chosen such that each term has a conformal weight $1$ and is thus invariant under the Weyl transformation
\begin{equation}
    \hat g = \e^{2 \omega} \hat g', \qquad
    \sigma = \sigma' - \omega, \qquad
    \hat{\mc O}_i = \e^{- 2 h_i \omega}\, \hat{\mc O}_i,
\end{equation}
which leads to the condition
\begin{equation}
    a_i (Q - a_i) + h_i = 1
\end{equation}
(the $a_i^2$ term comes from the regularization of the exponential).
The solution is
\begin{equation}
    a_i = \frac{Q}{2} - \sqrt{\frac{Q^2}{4} + h_i - 1}
\end{equation}
where the sign is found by matching to the semi-classical solution $Q \to \infty$
\begin{equation}
    a_i \sim \frac{1}{Q} (1 - h_i).
\end{equation}
The multiplication of the matter primary by a Liouville primary is called gravitational dressing.

A derivation of \eqref{eq:ddk-ansatz} has been suggested in~\cite[sec.~3]{Zamolodchikov:2002:ScalingLeeYangModel}.
Starting from the matter partition function
\begin{equation}
    Z_m[g]
        = \int \mc D \psi \, \e^{- S_{\text{cft}}[g, \psi] - S_{p}[g, \psi]}
        = Z_{\text{cft}}[g] \Mean{\exp\left(- \lambda \int \dd^2 \sigma\, \sqrt{g}\, \mc O \right)}_{g, \text{cft}},
\end{equation}
where $Z_{\text{cft}}[g]$ arises from the normalization of the correlation function, the exponential can be expanded perturbatively in the coupling constant
\begin{equation}
    \label{part:matter-expansion}
    \frac{Z_m[g]}{Z_{\text{cft}}[g]}
        = \sum_{n=0}^\infty \frac{(-\lambda)^n}{n!}
            \int \Mean{\mc O(x_1) \cdots \mc O(x_n)}_{g, \text{cft}} \prod_{i=1}^n \sqrt{g(x_i)}\, \dd^2 x_i.
\end{equation}
Since the correlation functions are computed in the CFT, one can use the relations \eqref{eq:weyl-g-O} in order to express the quantities in the conformal gauge
\begin{equation}
    \frac{Z_m[g]}{Z_{\text{cft}}[g]}
        = \sum_{n=0}^\infty \frac{(-\lambda)^n}{n!}
            \int \mean{\hat{\mc O}(x_1) \cdots \hat{\mc O}(x_n)}_{\hat g, \text{cft}} \prod_{i=1}^n \e^{2 (1 - h_i) \sigma(x_i)} \sqrt{\hat g(x_i)}\, \dd^2 x_i.
\end{equation}
The expansion can be resummed
\begin{equation}
    \frac{Z_m[g]}{Z_{\text{cft}}[g]}
        = \frac{1}{Z_{\text{cft}}[\hat g]} \int \dd_{\hat g} \psi\, \e^{- S_{\text{cft}}[\hat g, \psi] - S^{(p)}_{\text{DDK}}[\hat g, \sigma, \psi]}.
\end{equation}
Finally, \eqref{eq:ddk-ansatz} is recovered by using the relation
\begin{equation}
    \ln \frac{Z_{\text{cft}}[g]}{Z_{\text{cft}}[\hat g]} = Q^2 \, S_L[\hat g, \sigma].
\end{equation}

The semi-classical result can be recovered by a direct computation with the action
\begin{equation}
    S_p[g, \psi] = \sum_i \lambda^i \int \dd^2 x \sqrt{g} \, \mc O_i
        = \sum_i \lambda^i \int \dd^2 x \sqrt{\hat g} \, \e^{2 (1 - h_i) \sigma} \hat{\mc O}_i
        = S^{(p)}_{\text{DDK}}[\hat g, \sigma, \psi].
\end{equation}
Note that this does not follow from a direct integration of the quantum expectation value of the trace \eqref{eq:trace-pert}
\begin{equation}
    \mean{T^{(p)}}_g = 4\pi \sum_i \lambda^i \mean{\mc O_i}_g
        \sim 4\pi \sum_i \lambda^i \mc O_{i,\text{cl}}
\end{equation}
because one finds an additional factor of $(1 - h_i)^{-1}$.

\subsection{Discussion}
\label{sec:ddk:discussion}

As mentioned in the introduction of this section, arguments showing that it is not clear whether the DDK ansatz of the gravitational action for a deformed CFT is valid can already be found in the existing literature or derived from elementary facts:
\begin{enumerate}
    \item The action $S^{(p)}_{\text{DDK}}[g, \sigma, \psi]$ is not a Wess--Zumino action since it depends on both the matter fields and the Liouville mode.
    This means that the ansatz does not fit directly inside the conformal gauge formalism despite the appearance.

    To be more explicit, compare the DDK ansatz \eqref{eq:ddk-ansatz} with the conformal gauge action \eqref{eq:cg-action} where the gravitational action would be of the form
    \begin{equation}
        S_{\text{grav}}[\hat g, \sigma] = Q^2 S_L[\hat g, \sigma] + S^{(p)}_{\text{grav}}[\hat g, \sigma].
    \end{equation}
    The discrepancy between $S_{\text{DDK}}$ and $S_{\text{cg}}$ can be summarised by the fact that the non-conformal contribution to the gravitational $S^{(p)}_{\text{grav}}$ and matter $S_p$ actions have been replaced by a single term $S^{(p)}_{\text{DDK}}$.

    Moreover, the definition of $Q^2$ in \eqref{eq:ddk-ansatz} contains the central charge of the ghosts and of every matter sector, which means that the different CFT sectors are not decoupled anymore.
    The conformal gauge action \eqref{eq:cg-action} does not show any sign of such coupling.

    \item From the previous point, it follows that the DDK action cannot be recovered by integrating the trace anomaly or by computing the ratio of the partition functions in two different metrics.

    \item The ansatz was proposed by requiring the total action to be conformally invariant term by term (whereas the validity of the conformal gauge approach asks only for the invariance of the total action).
    In the DDK approach, this translates into the fact that the gravitationally dressed deformations should be exactly marginal operators.
    However, this is generically not the case beyond the leading order in the coupling~\cite{Klebanov:1993:GravitationalDressingRenormalization, Schmidhuber:1993:ExactlyMarginalOperators, Ambjorn:1994:2DQuantumGravity, Tanii:1994:PhysicalScalingRenormalization, Dorn:1995:GravitationalDressingRenormalization, Schmidhuber:1995:RGFlowRandom, Dorn:1996:SchemeDependenceGravitational}.
    This can be understood from the fact that the presence of the operators inside the action modifies the renormalization conditions and thus one cannot expect a tree-level condition to hold beyond the semi-classical level.
    This indicates that the DDK ansatz, beyond the first order approximation, should be supplemented with an infinite number of terms~\cite{Dorn:1995:GravitationalDressingRenormalization}.
    This may lead to a well-defined gravitational action, but computing the higher-order corrections is a formidable task and not more easy than a direct computation of the gravitational action in the conformal gauge (in particular for weights far from $1$).

    Let us illustrate this at the second order.
    Given the OPE coefficients $C_{ij}^k$
    \begin{equation}
        V_i(x) V_j(y) \sim C_{ij}^k \abs{x - y}^{2 (h_k - h_i - h_j)} V_k(y)
    \end{equation}
    the beta functions for operators with $h_i \sim 1$ and for vanishing cosmological constant are
    \begin{equation}
        \beta^i = (\Delta_i^j - 2 \, \delta_i^j) \lambda^j + \pi \, C_{jk}^i \, \lambda^j \lambda^k + O(\lambda^3)
    \end{equation}
    where $\Delta_i^j$ is the normalization of the two-point function (the $O(\lambda^3)$ term has been computed in~\cite{Dorn:1995:GravitationalDressingRenormalization, Schmidhuber:1995:RGFlowRandom}).
    One can add a term
    \begin{equation}
        S^{(p,2)}_{\text{DDK}} = \frac{\pi}{Q (1 + 4 a_k)} \, C_{ij}^k \lambda^i \lambda^j \int \dd^2 \sqrt{\hat g} \, \sigma \, \e^{2 a_k Q \sigma} \hat{\mc O}_k(\psi)
    \end{equation}
    to ensure that the beta functions vanish at quadratic order (in practice this implies that $\Delta_i^j$ becomes coupling-dependent).
    Note that, without this additional term, the vanishing of the beta functions at $O(\lambda)$ is obtained thanks to the gravitational dressing.

    \item The derivation from~\cite{Zamolodchikov:2002:ScalingLeeYangModel} presented in the previous section may yield an incorrect result because of the formal manipulations in the functional integral.
    There are three possible sources of errors: 1) the correlation functions have logarithmic singularities~\cite[p.~4]{Ambjorn:1994:2DQuantumGravity} which are not regularized, 2) the functional integral and infinite sum are exchanged (twice), 3) the infinite series (with divergent terms) is directly resummed.
    A rigorous computation would require to regularize the correlation functions, for example by introducing a connection on the CFT space~\cite{Kutasov:1989:GeometrySpaceConformal, Sonoda:1993:ConnectionTheorySpace, Ranganathan:1993:NearbyCFTsOperator, Ranganathan:1994:ConnectionsStateSpaceConformal}.
\end{enumerate}

Since the massive Majorana theory is a CFT with $c = 1/2$ deformed by a primary operator
\begin{equation}
    S_{\text{cft}} = \frac{1}{4\pi} \int \dd^2 x\, \sqrt{g} \, \bar\Psi i \slashed\grad \Psi,
    \qquad
    S_p[g, \psi] = \frac{m}{4\pi} \int \dd^2 x\, \sqrt{g} \, \bar\Psi \gamma_* \Psi,
\end{equation}
the current paper provides an explicit example to test the DDK ansatz.
The conformal weights of the holomorphic and anti-holomorphic components are respectively $(h, \bar h) = (1/2, 0)$ and $(h, \bar h) = (0, 1/2)$, such that the weight of the energy $\bar\Psi \gamma_* \Psi$ is $(h, \bar h) = (1/2, 1/2)$.
Then, the DDK ansatz gives
\begin{equation}
    \begin{gathered}
    S_{\text{DDK}}[g, \sigma, \psi] = - Q^2 \, S_L[\hat g, \sigma] + S_{\text{cft}}[\hat g, \psi] + S^{(p)}_{\text{DDK}}[\hat g, \sigma, \psi],
    \\
    S^{(p)}_{\text{DDK}}[\hat g, \sigma, \psi] = \frac{m}{4\pi} \int \dd^2 x\, \sqrt{g} \, \e^{2 a Q \sigma} \bar\Psi \gamma_* \Psi,
    \end{gathered}
\end{equation}
We do not provide the $Q^2$ and $a$ because their values are model dependent (for example, if one adds other matter sectors).
This action looks different from \eqref{expression-Sgrav-final}, which we obtained by the rigorous heat kernel method.
In particular, the DDK action \eqref{eq:ddk-action} is linear in the coupling constant while the gravitational action for the massive fermion is not.
The action \eqref{expression-Sgrav-final} is by construction consistent with the conformal gauge, see \eqref{eq:cg-action} and none of the points listed in the list applies to it.
According to \eqref{Sgrav-small-mass}, the first-order correction will be proportional to $m^2$ instead of $m$ for the DDK ansatz.

This can be summarized by saying that a \emph{deformation of the matter CFT} does not seem to be equivalent to a \emph{deformation of the Liouville plus matter CFT}, at least beyond the semi-classical approximation.\footnotemark{}
\footnotetext{%
    See also~\cite[pp.~19--20]{Martinec:2014:ModelingQuantumGravity} for related discussions.
}%
A possible explanation is that symmetries are not sufficient to determine the form of the action when locality is lost, which is certainly the case in a theory of gravity with massive matter (for example, the Mabuchi action is non-local when expressed in terms of the Liouville field).

It would be interesting to establish our result using CFT methods, for example by performing the computations from~\cite{Zamolodchikov:2002:ScalingLeeYangModel} using the tools from~\cite{Kutasov:1989:GeometrySpaceConformal, Sonoda:1993:ConnectionTheorySpace, Ranganathan:1993:NearbyCFTsOperator, Ranganathan:1994:ConnectionsStateSpaceConformal}.

Finally, note that these problems might just be apparent and result from a “bad” parametrization of the action and functional integrals.
This would amount to prove that
\begin{equation}
    \int \mc D \sigma \, \mc D \psi \, \e^{- S_{\text{cft}}[\hat g, \psi] - S^{(p)}_{\text{DDK}}[\hat g, \sigma, \psi]}
        = \int \mc D \sigma \, \mc D \psi \, \e^{- S_{\text{cft}}[\hat g, \psi] - S_p[\hat g, \psi] - S^{(p)}_{\text{grav}}[\hat g, \psi]}.
\end{equation}
It is also possible that such a relation would hold only after incorporating all the corrections to the DDK action needed to make the beta functions vanish.
While establishing directly this identity might be difficult, one may also prove it either by showing that the correlation functions computed from both sides satisfy the same Ward identities, or by computing numerically the correlation functions and showing that they agree.

Different interpretations can be given to the DDK ansatz, which is reflected by its various uses in the literature.
The previous discussion applies \emph{only} when it is used to postulate an action for a deformed CFT coupled to gravity.
These comments do not apply when the DDK action is used by itself as a model (for example of a statistical system); instances of such approaches (even if there is a confusion with the previous case) include~\cite{David:1988:ConformalFieldTheories, Eguchi:1993:C1LiouvilleTheory, Zamolodchikov:2002:ScalingLeeYangModel, Doyon:2004:IsingFieldTheory, Zamolodchikov:2006:MassiveMajoranaFermion, daCunha:2009:CrumpledWiresLiouville}.
It should be clear that this approach gives up the link with gravity and study the action and the associated functional integral just for themselves: as a consequence, any constraint originating from gravity can be relaxed (in particular the Wess--Zumino form and the marginal scaling conditions) and our comments do not apply.
The fact that it defines a well-defined statistical system can also be motivated from the mapping to matrix models~\cites{Zamolodchikov:2005:PerturbedConformalField, Zamolodchikov:2006:MassiveMajoranaFermion}[sec.~4]{Schmidhuber:1993:ExactlyMarginalOperators}.
Similarly, the gravitational dressing of operators in correlation functions such as
\begin{equation}
    \Mean{\prod_i \e^{2 a_i Q \sigma(x_i)} \mc O_i(x_i)}
\end{equation}
do not pose any problems since the scaling dimensions are not modified in this case (for a discussion see~\cite{Distler:1989:ConformalFieldTheory, Dorn:1989:ScalingDimensionsVertex, DHoker:1990:2DQuantumGravity, Ginsparg:1991:MatrixModels2d, Ferrari:2012:GravitationalActionsTwo}).

\section{Proof of various identities}
\label{app:proof-identities-Kzeta}

The goal of this section is to express $\int \dd^2 x \sqrt g \, \delta \sigma(x) \tr_D \zeta(2, x, x)$ in terms of the heat kernel, i.e.\ to prove \eqref{eq:Kzeta2}.
The variation of $K(t)$ is given by
\begin{align}
	\delta K(t)
		= \delta \tilde K(t)
		&
		= - t \sum_{n \neq 0} \e^{-\Lambda_n t} \delta \Lambda_n \nonumber
		\\ &
		= - 2 t \left(\frac{\dd}{\dd t} + m^2\right) \int \dd^2 x \sqrt g \, \delta \sigma(x) \tr_D \tilde K(t, x, x) \nonumber \\
			& \qquad
			- t \sum_{n \neq 0} \e^{-\Lambda_n t} \int \dd^2 x \sqrt g \, \Psi_n^\dagger(x) \partial_\nu (\delta \sigma) \gamma^{\mu\nu} \nabla_\mu \Psi_n(x) \nonumber \\
			& \qquad
			- \frac{t}{2} \int \dd^2 x \sqrt g \, \Delta(\delta \sigma(x)) \tr_D \tilde K(t, x, x).
\end{align}
To compute the second term, we introduce the generalized heat kernel
\begin{equation}
	K(s, t, x, y)
		= \sum_n \e^{-\Lambda_n t} \frac{\Psi_n(x) \Psi_n(y)}{\Lambda_n^s}.
\end{equation}
In particular
\begin{equation}
	K(0, t, x, y)
		= K(t, x, y).
\end{equation}
We can then redo the computations we have done for the variation of the zeta function.
For instance, the computation of $I_3$ in \eqref{eq:I-values} can be generalized as
\begin{equation}
	\begin{multlined}
	\sum_{n \neq 0} \e^{-\Lambda_n t} \int \dd^2 x \sqrt g \, \frac{\partial_\nu (\delta \sigma) \Psi_n^\dagger(x) \gamma^{\mu\nu} \nabla_\mu \Psi_n(x)}{\Lambda_n^{s+1}}
		\\
		= - \frac{1}{2} \int \dd^2 x \sqrt g \, \delta \sigma(x) \tr_D (\Delta \tilde K(s+1, t, x, x)).
	\end{multlined}
\end{equation}
Taking $s = -1$, we find that
\begin{equation}
	\begin{multlined}
	\sum_{n \neq 0} \e^{-\Lambda_n t} \int \dd^2 x \sqrt g \, \partial_\nu (\delta \sigma) \Psi_n^\dagger(x) \gamma^{\mu\nu} \nabla_\mu \Psi_n(x)
		\\
		= - \frac{1}{2} \int \dd^2 x \sqrt g \, \delta \sigma(x) \tr_D (\Delta \tilde K(t, x, x)).
	\end{multlined}
\end{equation}
We then have
\begin{equation}
	\delta K(t)
		= - 2t \left(\frac{\dd}{\dd t} + m^2\right) \int \dd^2 x \sqrt g \, \delta \sigma(x) \tr_D \tilde K(t, x, x).
\end{equation}
Note that the differential equations satisfied by $K$ and $K^{(0)}$ imply that
\begin{equation}
	K(t, x, y)
		= \e^{-m^2 t} K^{(0)}(t, x, y).
\end{equation}
The eigenfunctions expansions show that the same is true between $\tilde K$ and $\tilde K^{(0)}$, which implies
\begin{equation}
	\delta K(t)
		= - 2t \e^{-m^2 t} \frac{\dd}{\dd t} \int \dd^2 x \sqrt g \, \delta \sigma(x) \tr_D \tilde K^{(0)}(t, x, x).
\end{equation}
We then have
\begin{align}
	\int_0^\infty \dd t \, \frac{\e^{m^2 t} - m^2 t -1}{t} \delta \tilde K(t) &
		= -2 \left[\int \dd^2 x \sqrt g \, \delta \sigma \tr_D \tilde K^{(0)}(t, x, x)\right]_0^{+\infty}
			\nonumber
			\\ & \quad
			+ 2 \int_0^{+\infty} \dd t \, (m^2 t + 1)\left(m^2 + \frac{\dd}{\dd t}\right) \int \dd^2 x \sqrt g \, \delta \sigma \tr_D \tilde K(t, x, x)
			\nonumber
		\\ &
		= -2 \int \dd^2 x \sqrt g \, \delta \sigma \tr_D \left[\tilde K^{(0)}(t, x, x) - \tilde K(t, x, x) \right]_0^{+\infty}
		\nonumber
			\\ & \qquad
			+ 2 m^2 \int \dd t \, \int \dd^2 x \sqrt g \, \delta \sigma \tr_D \tilde K(t, x, x)
			\nonumber
			\\ & \qquad
			+ 2 m^2 \int \dd t \, t \, \frac{\dd}{\dd t} \int \dd^2 x \sqrt g \, \delta \sigma \tr_D \tilde K(t, x, x)
			\nonumber
			\\ & \qquad
			+ 2 m^4 \int \dd t \, t \int \dd^2 x \sqrt g \, \delta \sigma \tr_D \tilde K(t, x, x)
			\nonumber
	\\ &
		= 4 m^4 \int \dd^2 x \sqrt g \, \delta \sigma(x) \tr_D \tilde \zeta(2, x, x)
		\nonumber
		\\ &
		= 4 m^4 \int \dd^2 x \sqrt g \, \delta \sigma(x) \tr_D \zeta(2, x, x)
		\nonumber
		\\ & \qquad
			- 4 \int \dd^2 x \sqrt g \, \delta \sigma(x) \sum_i \Psi_{0,i}^\dagger(x) \Psi_{0, i}(x).
\end{align}

\printbibliography[heading=bibintoc]

\end{document}